\documentclass{aastex62}

\usepackage{lineno}
\usepackage{epsfig, graphicx, amsmath, amsfonts, amssymb, mathrsfs, textcomp}
\usepackage{subfigure}

\newcommand{\tw}{$^{12}$CO }
\newcommand{\tht}{$^{13}$CO } 
\newcommand{\ei}{C$^{18}$O }
\newcommand{\msun}{M$_\odot$}

\shorttitle{Sample article}
\shortauthors{Guo et al.}

\begin{document}

\title{A wide-field CO survey towards the California Molecular Filament}

\correspondingauthor{Weihua Guo, Xuepeng Chen}
\email{whguo@pmo.ac.cn, xpchen@pmo.ac.cn}

\author{Weihua Guo}
\affiliation{Purple Mountain Observatory and Key Laboratory of Radio Astronomy, Chinese Academy of Sciences,
10 Yuanhua Road, Nanjing 210034, China}
\affiliation{School of Astronomy and Space Science, University of Science and Technology of China, Hefei, Anhui 230026, China}

\author{Xuepeng Chen}
\affiliation{Purple Mountain Observatory and Key Laboratory of Radio Astronomy, Chinese Academy of Sciences,
10 Yuanhua Road, Nanjing 210034, China}
\affiliation{School of Astronomy and Space Science, University of Science and Technology of China, Hefei, Anhui 230026, China}

\author{Jiancheng Feng}
\affiliation{Purple Mountain Observatory and Key Laboratory of Radio Astronomy, Chinese Academy of Sciences,
10 Yuanhua Road, Nanjing 210034, China}
\affiliation{School of Astronomy and Space Science, University of Science and Technology of China, Hefei, Anhui 230026, China}

\author{Li Sun}
\affiliation{Purple Mountain Observatory and Key Laboratory of Radio Astronomy, Chinese Academy of Sciences,
10 Yuanhua Road, Nanjing 210034, China}
\affiliation{School of Astronomy and Space Science, University of Science and Technology of China, Hefei, Anhui 230026, China}

\author{Chen Wang}
\affiliation{Purple Mountain Observatory and Key Laboratory of Radio Astronomy, Chinese Academy of Sciences,
10 Yuanhua Road, Nanjing 210034, China}

\author{Yang Su}
\affiliation{Purple Mountain Observatory and Key Laboratory of Radio Astronomy, Chinese Academy of Sciences,
10 Yuanhua Road, Nanjing 210034, China}
\affiliation{School of Astronomy and Space Science, University of Science and Technology of China, Hefei, Anhui 230026, China}

\author{Yan Sun}
\affiliation{Purple Mountain Observatory and Key Laboratory of Radio Astronomy, Chinese Academy of Sciences,
10 Yuanhua Road, Nanjing 210034, China}
\affiliation{School of Astronomy and Space Science, University of Science and Technology of China, Hefei, Anhui 230026, China}

\author{Yiping Ao}
\affiliation{Purple Mountain Observatory and Key Laboratory of Radio Astronomy, Chinese Academy of Sciences,
10 Yuanhua Road, Nanjing 210034, China}

\author{Shaobo Zhang}
\affiliation{Purple Mountain Observatory and Key Laboratory of Radio Astronomy, Chinese Academy of Sciences,
10 Yuanhua Road, Nanjing 210034, China}

\author{Xin Zhou}
\affiliation{Purple Mountain Observatory and Key Laboratory of Radio Astronomy, Chinese Academy of Sciences,
10 Yuanhua Road, Nanjing 210034, China}

\author{Lixia Yuan}
\affiliation{Purple Mountain Observatory and Key Laboratory of Radio Astronomy, Chinese Academy of Sciences,
10 Yuanhua Road, Nanjing 210034, China}

\author{Ji Yang}
\affiliation{Purple Mountain Observatory and Key Laboratory of Radio Astronomy, Chinese Academy of Sciences,
10 Yuanhua Road, Nanjing 210034, China}
\affiliation{School of Astronomy and Space Science, University of Science and Technology of China, Hefei, Anhui 230026, China}

\begin{abstract}
	
	We present the survey of $^{12}$CO/$^{13}$CO/C$^{18}$O (J=1-0) toward 
	the California Molecular Cloud (CMC) within the region of 
	161.75$\degr \leqslant l \leqslant$ 167.75$\degr$, -9.5$\degr \leqslant b \leqslant$ -7.5$\degr$, using 
	the Purple Mountain Observatory (PMO) 13.7 m millimeter telescope. 
	Adopting a distance of 470\,pc, the mass of the observed molecular cloud estimated from $^{12}$CO, $^{13}$CO, and C$^{18}$O 
	is about 2.59$\times$10$^{4}$\,M$_\odot$, 0.85$\times$10$^{4}$\,M$_\odot$, and 0.09$\times$10$^{4}$\,M$_\odot$, respectively. 
	A large-scale continuous filament extending about 72 pc  is revealed from the $^{13}$CO images.
	A systematic velocity gradient perpendicular to the major axis appears and is measured to be $\sim$ 0.82 km\,s$^{-1}$\,pc$^{-1}$.
	The kinematics along the filament shows an oscillation pattern with a fragmentation wavelength of $\sim$ 2.3 pc and velocity amplitude 
	of $\sim 0.92 \mathrm{~km} \mathrm{~s}^{-1}$, which may be related with core-forming flows.
	Furthermore, assuming an inclination angle to the plane of the sky of $45^{\circ}$, 
	the estimated average accretion rate is $\sim$ 101 M$_\odot$\,Myr$^{-1}$ for the cluster  
	LkH$\alpha$ 101 and $\sim$ 21 M$_\odot$\,Myr$^{-1}$ for the other regions.
	In the C$^{18}$O observations, the large-scale filament could be resolved into multiple substructures and 
	their dynamics are consistent with the scenario of filament formation from converging flows.
	Approximately 225 C$^{18}$O cores are extracted, of which 181 are starless cores. 
	Roughly $37\,\%\, (67/181)$ of the starless cores have $\alpha_{\text{vir}}$ less than 1.
        Twenty outflow candidates are identified along the filament.
	Our results indicate active early-phase star formation along the large-scale filament in the CMC region.
	
\end{abstract}

\keywords{clouds -- ISM:  structure -- ISM: kinematic and dynamic -- ISM: outflows -- stars: formation}

\section{Introduction} \label{sec:intro}
	 	
	Filamentary structures are ubiquitous in the interstellar medium (ISM). 
	In the last decade, based on multi-wavelength observations (e.g., surveys with the $Herschel$ Space Observatory), 
	great progresses have been made 
	in the relationship between filamentary molecular clouds and star formation \citep{2010A&A...518L.102A,2010A&A...518L.100M}. 
	It is widely accepted that filamentary molecular clouds play an important role in the process of star formation.	
	Both observational and theoretical studies suggest that large-scale supersonic compressed gas forms reticular structures firstly 
	in the ISM, and then reticular structures fragment and form prestellar cores due to the force of self-gravity, 
	finally the cores evolve into stars \citep{2014prpl.conf...27A}.
	Velocity gradients perpendicular \citep{2018ApJ...853..169D,2019A&A...623A..16S} and along \citep{2011A&A...533A..34H,2018ApJ...855....9L} 
	filaments have been found in favor of this picture.
	However, it is necessary to survey and investigate more regions to complement the theory about the 
	filament formation and star formation.
	
\begin{figure}[htb]
		\centering
		\includegraphics[width=1.0\textwidth]{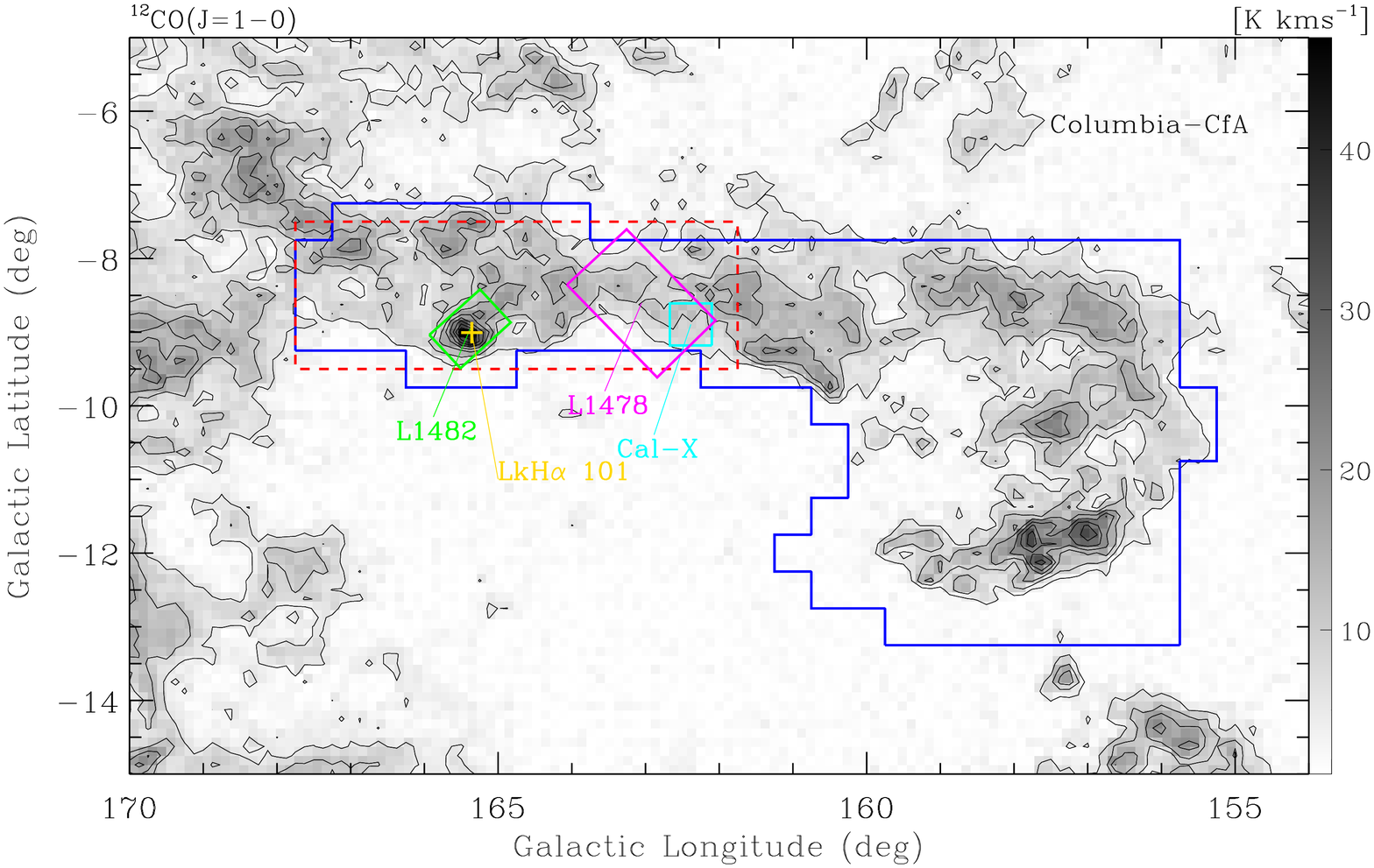}%\textit{(a)}
		\caption{Integrated intensity map of the Columbia-CfA \tw(J=1-0) survey towards the Galactic Plane within 
		154$\degr$ $\leqslant l \leqslant$ 170$\degr$, -15$\degr$ $\leqslant b \leqslant$ -5$\degr$ \cite[from] []{2001ApJ...547..792D}. 
		The integrated velocity range is from -15 to 10 km\,s$^{-1}$. 
		The contours are from 10\,\% to 90\,\% by a step of 10\,\% of the maximum intensity value ($\sim$ 46 K km\,s$^{-1}$). 
		The blue polygon marks the scanning coverage of the MWISP CO survey (Wang et al. in prep).
		The red dashed box indicates the region focused in this work, based on the MWISP data.  
		The green, magenta and cyan polygons show the observed regions toward 
		L1482 \citep{2014A&A...567A..10L}, L1478 \citep{2019ApJ...877..114C}, and Cal-X \citep{2017ApJ...840..119I}, respectively.
		The plus sign represents the location of the $\mathrm{LkH}\alpha$ 101 cluster throughout all the figures in this work.
		}
		\label{fig-m0_cfa}
	\end{figure}

	Located in the Taurus-Perseus-Auriga complex \citep{1987ApJS...63..645U, 2001ApJ...547..792D},  the CMC
	is identified to be a giant molecular cloud (GMC), which is similar to the Orion Molecular Cloud (OMC) in mass 
	($\sim$ 10$^{5}$\,\msun) and size ($\sim$ 80 pc), but in an earlier evolution stage \citep{2009ApJ...703...52L,2018ApJ...852...73B}. 
	%with much less star formation \citep{2009ApJ...703...52L,2018ApJ...852...73B}. 
	Compared with the OMC, the CMC was found to contain 15$-$20 times fewer young stellar objects (YSOs) in number \citep{2014ApJ...786...37B}.
	Only one cluster LkH$\alpha$ 101 was found to be embedded in the reflection nebula NGC 1579  in the CMC \citep{2008hsf1.book..390A},
	and the estimated age of the cluster is $\sim$ 1 Myr \citep{2010ApJ...715..671W}.
	The Columbia-CfA 1.2\,m \tw(J=1-0) survey showed that the CMC is an elongated molecular cloud (see Figure \ref{fig-m0_cfa}). 
	The $Herschel$ observations further showed filamentary structures in the CMC \citep{2013ApJ...764..133H, 2017A&A...606A.100L}, 
	covering an area roughly from 160$\degr$ to 168$\degr$ along the Galactic Longitude and 
	from -10$\degr$ to 7$\degr$ along the Galactic Latitude.
	Therefore, the CMC can serve as an ideal laboratory to investigate the structures and dynamics of large-scale filamentary gas structures
	 and early-phase  star formation therein. 
	 
	In recent years, various observations have been carried out toward subregions  in the CMC, such as L1482, L1478 and California-X.
	\cite{2014A&A...567A..10L} investigated the kinematic and dynamical states of the L1482 nebula, the most active star formation region in the CMC,  
	and investigated the physical properties of dense clumps by molecular line emission observations of the \tw (2-1), 
	\tw (3-2), and \tht (1-0).
	\cite{2017ApJ...840..119I} studied the relationship between filaments, cores, and outflows in the California-X region 
	centered at (l, b)=(162.4$\degr$, -8.8$\degr$) by \tw (2-1) and \tht (2-1). 
	%Physical properties of filaments, cores and outflow included in the Cal-X structure are investigated using \tw(2-1) and \tht(2-1) in 2017.
	\cite{2017A&A...606A.100L} provided a census of YSOs by combining the data from $IRAS$ \citep{2009ApJ...703...52L},  
	$Herschel$ \citep{2013ApJ...764..133H} and $Spitzer$ \citep{2014ApJ...786...37B}.
	\cite{2018A&A...620A.163Z} performed  single-pointing IRAM-30\,m observations of molecular lines near 90 GHz to the CMC and 
	derived the physical properties of dense cores extracted from the $Herschel$ data.
	\cite{2018ApJ...852...73B} identified a sample of protostellar objects by the 850 $\mu$m and 450 $\mu$m observations using the 
	SCUBA-2 as part of the JCMT Gould Belt Legacy Survey.
	\cite{2019ApJ...877..114C} performed observations of molecular lines in the frequency range of 85$-$115 GHz to study the kinematics  of the 
	filaments and chemical evolution of the dense cores toward the L1478 region. 
	However, the studies mentioned above mainly focused on the small filamentary structures and dynamic properties in the subregions of the CMC, 
	and there is currently a lack of study on the structure and dynamics of this molecular cloud  as a whole.

	The Milky Way Imaging Scroll Painting (MWISP) project\footnote{\url{http://www.radioast.nsdc.cn/mwisp.php}} is a large CO\,(J=1--0) 
	multi-line survey towards the northern Galactic Plane, using the PMO-13.7 m millimeter telescope. 
	This unbiased survey provides us a good opportunity to 
	study large-scale filamentary molecular clouds, as well as star formation therein \cite[see, e.g.][]{2019ApJ...880...88X}. 
	As a part of the MWISP survey, we present in this work the results of the three CO (J=1-0) isotopologue line observations toward the CMC.
	In Section 2, we describe the CO line observations and data reduction. 
	We present observational results in Section 3 and discuss these results in Section 4.
	In Section 5, we make a summary of these discussions.

\section{Observations and Data Reduction} \label{sec-data}
	
	The three CO isotopologue lines observations were simultaneously 
	performed from February 2012 to May 2014 with the PMO-13.7 m millimeter telescope located in Delingha, China.
	The observations were performed using a 3$\times$3 beams sideband separation Superconducting Spectroscopic Array Receiver (SSAR) system \citep{2012ITTST...2..593S}. 
	The frequency of the local oscillator (LO) was carefully selected, and thus the
	${ }^{12} \mathrm{CO}$ is at the upper sideband (USB), and ${ }^{13} \mathrm{CO}$ and $\mathrm{C}^{18} \mathrm{O}$ 
	 are at the lower sideband (LSB), respectively. 
	The pointing uncertainty is approximately within 5$\arcsec$. 
	The observations were conducted by position-switch On-The-Fly (OTF) mode, and the telescope scaned the sky 
	along both the longitude and latitude directions at a constant rate of 50$\arcsec$s$^{-1}$.
	Velocities were all given with respect to the local standard of rest (LSR). 
	The observational parameters of frequency, half-power beamwidth (HPBW), velocity resolution 
	($\Delta \mathrm{v}$), rms noise levels ($\sigma_\mathrm{rms}$), 
	typical system temperatures, and beam efficiencies ($\eta_{\rm MB}$) are listed in Table \ref{tb-obser}.
	For more information about the telescope, to refer to the 
	status report of the PMO-13.7 m millimeter telescope\footnote{\url{http://www.radioast.csdb.cn/zhuangtaibaogao.php}}. 
	All data were reduced by GILDAS/CLASS package\footnote{\url{http://www.iram.fr/IRAMFR/GILDAS}}. 
	After removing bad channels and abnormal spectra and correcting the first order (linear) baseline fitting, 
	the data were regridded into the standard FITS files  with a pixel size of 30$\arcsec \times$ 30$\arcsec$ 
	(approximately half of the beam size). 
		
	\begin{deluxetable}{lcccccc}
		\tabletypesize{\small}
		\setlength{\tabcolsep}{0.15in}
		\tablecaption{PMO-13.7 m observational parameters \label{tb-obser}}
		\tablewidth{0pt} 
		%\tablenum{5}
		\tablehead{ \colhead{Tracer} & \colhead{Frequency} & \colhead{HPBM} & \colhead{$\Delta \mathrm{v}$} & \colhead{Mean Noise ($\sigma_\mathrm{rms}$)} & \colhead{System Temperatures} &\colhead{$\eta_{\rm MB}$}\\
		\colhead{}&\colhead{(GHz)} &\colhead{($\arcsec$)} & \colhead{(kms$^{-1}$)} & \colhead{(K)} & \colhead{(K)} & \colhead{(K)}
		} 
		\colnumbers
		\startdata 
		$^{12}$CO(J=1-0)  & 115.271  & 52 &  0.16     & 0.48   & $\sim$ 270     & 44\%     \\
		$^{13}$CO(J=1-0)  & 110.201  & 55 &  0.17     & 0.28   & $\sim$ 175     & 48\%     \\
		C$^{18}$O(J=1-0)  & 109.782  & 55 &  0.17     & 0.28   & $\sim$ 175     & 48\%     \\
		\enddata
	\end{deluxetable}

\section{Results} \label{sec:result}
	  
\subsection{Overview of the CMC} \label{sec:overview}
        	
	As shown in Figure \ref{fig-m0_cfa},  the blue solid polygon marks the  whole CMC area observed by the 
	MWISP CO survey, which covers 37.5 square degree. 
	In this work, we focus on the 6$\degr \times$ 2$\degr$ region 
	(161.75$\degr \leqslant l \leqslant$ 167.75$\degr$, -9.5$\degr \leqslant b \leqslant$ -7.5$\degr$) 
	as outlined by the red dashed box, which contains the filamentary structure suggested by the $Herschel$ 
	observations \cite[see, e.g,][]{2013ApJ...764..133H}. 
	The MWISP CO data towards the larger-scale observed region will be presented in other work (Wang et al. in prep).
		
\begin{figure}[htb]
	\centering
		\includegraphics[width=1\textwidth]{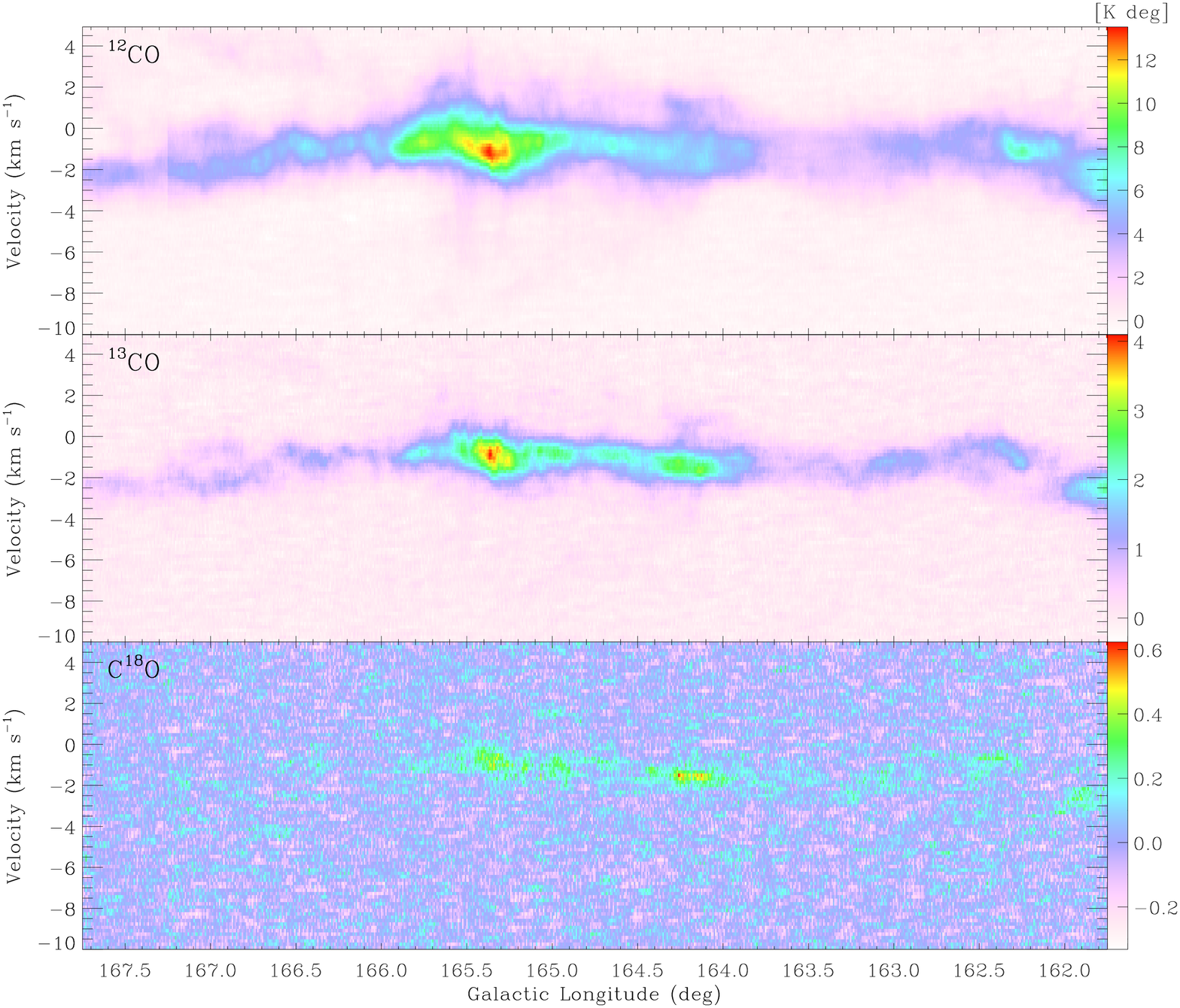} %\textit{(a)}
		\caption{The LV maps of  \tw (upper panel), \tht (middle panel), and \ei (bottom panel) 
                toward the Galactic Plane within 161.75$\degr \leqslant l \leqslant$ 167.75$\degr$ 
                and -10 kms$^{-1} \leqslant v \leqslant$ 5 km\,s$^{-1}$. 
                The Latitude is integrated from -9.5$\degr$ to -7.5$\degr$. 
                 }
		\label{fig-lv}
	\end{figure}

\begin{figure}[htb]
	\centering
		\includegraphics[width=1\textwidth]{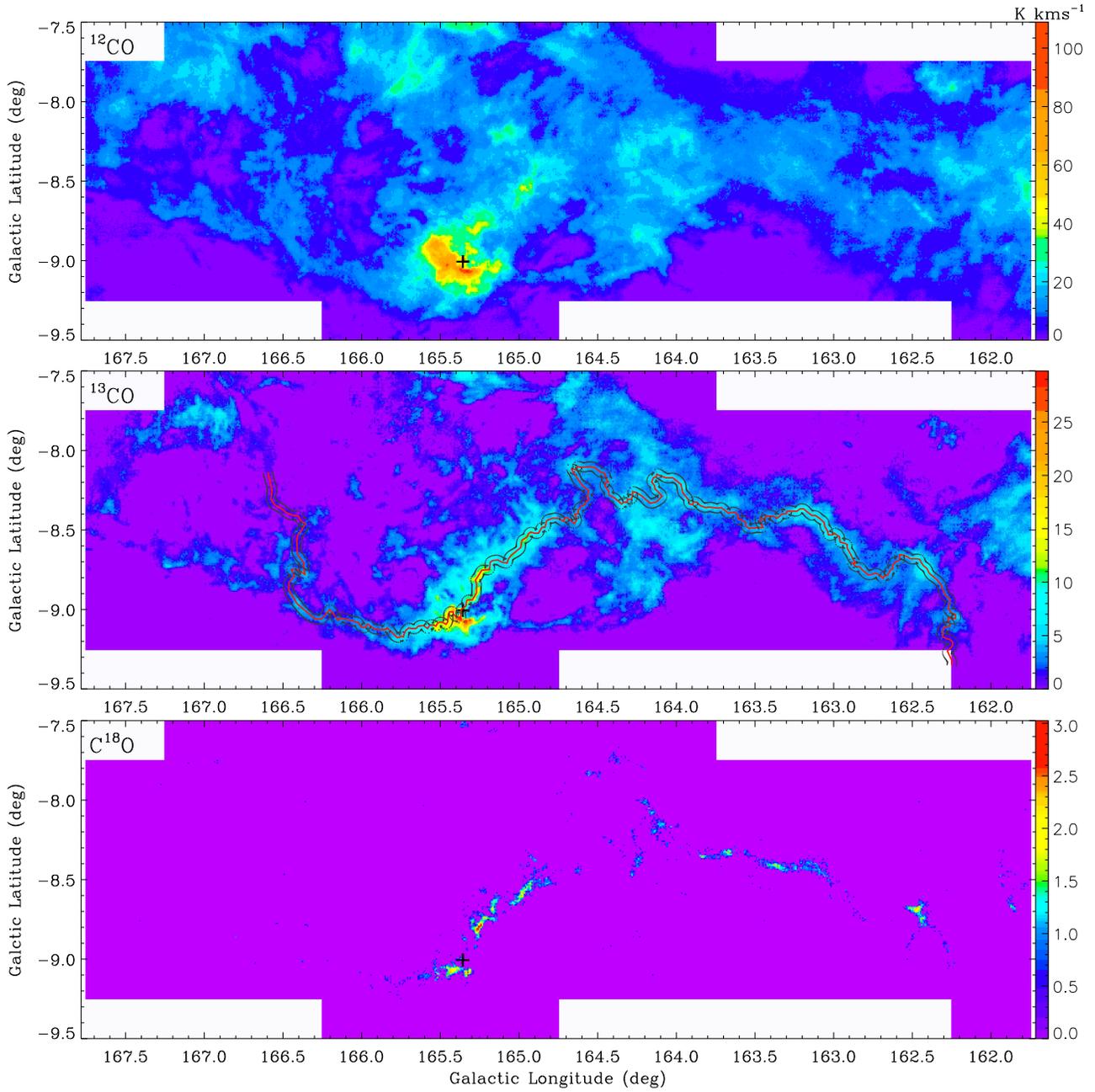} %\textit{(a)}
		\caption{The panel from top to bottom is the m0 maps of $^{12}$CO, $^{13}$CO, and C$^{18}$O integrated 
		in  [-10, 5]\,km\,s$^{-1}$, respectively. 
		The unit of the colorbar is K km\,s$^{-1}$.  
		The red curve in the middle panel indicate the ridgeline of the CMF with width (black dotted lines) of 
		$0.6\,\mathrm{pc}$ $(\sim 8\,$ pixels$)$
		}
		\label{fig-m0}
	\end{figure}

\begin{figure}[htb]
		\centering
		\includegraphics[width=1.0\textwidth]{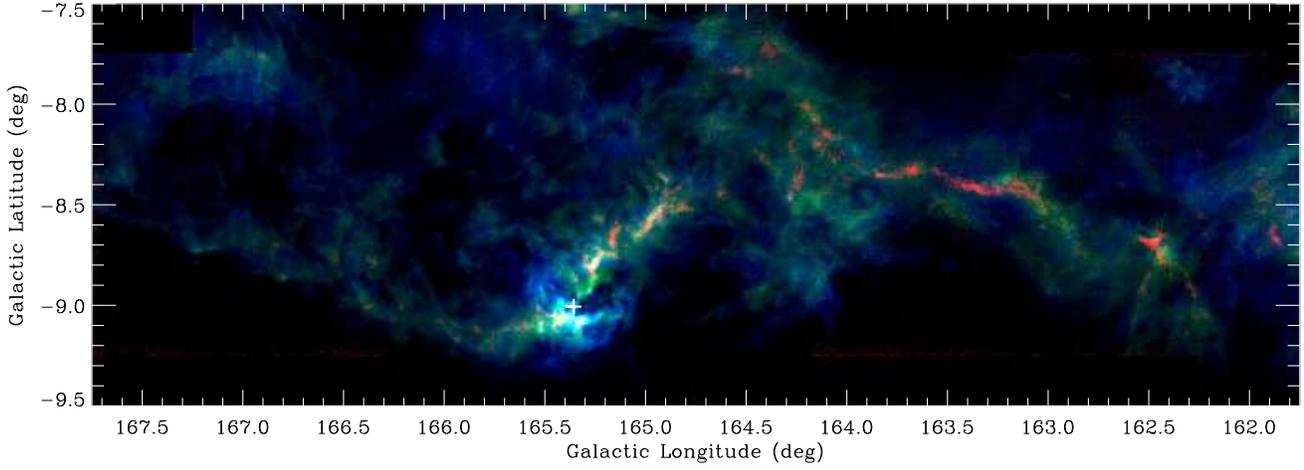}%\textit{(a)}
		\caption{The three-color image of the \tw(blue), \tht(green), and \ei(red) emission integrated in  [-10, 5] km\,s$^{-1}$. 
		The intensity scales are set to be [0, 50] K km\,s$^{-1}$ for $^{12}$CO,  [0, 15] K km\,s$^{-1}$ for $^{13}$CO, 
               and [0, 1.5] K km\,s$^{-1}$ for C$^{18}$O, respectively.}
		\label{fig-m0_thrcolor}
	\end{figure}

	In the Columbia-CfA \tw(J=1-0) survey, the velocity distribution of the CMC is mainly in the range 
	of $\sim$ [-15, 10] km\,s$^{-1}$ \cite[see, e.g,][]{2009ApJ...703...52L}. 
	Figure \ref{fig-lv} shows the Longitude-Velocity (LV) diagram of the MWISP \tw emission,
	the main velocity component is coherent and narrowly distributed within [-10, 5] km\,s$^{-1}$. 
	 It is found empirically that the typical gas densities traced by $^{12}$CO, $^{13}$CO, and C$^{18}$O are 
	 $\sim$ $10^{2}$ cm$^{-3}$, $\sim$ $10^{3}$ cm$^{-3}$, and $\sim$ $10^{4}$ cm$^{-3}$, respectively \citep{2005ApJ...634..476Y}.
	As shown in Figure \ref{fig-m0},  
	the \tw emission distributes diffusely and peripherally in the integral intensity (m0) map integrated in [-10, 5] km\,s$^{-1}$. 
	The \tht emission shows the large-scale filamentary structures extending about 5$\degr$ from east to 
	west\footnote{All the directions mentioned in this work refer to the direction in the Galactic coordinate system.}. 
	\ei is generally found to trace dense cores. %at a spital resolution of $\sim$ 1$\arcmin$
	In this work, bright \ei emission also outlines the ridgeline of the filamentary structure in the CMC region. 
	As shown in Figure \ref{fig-m0_thrcolor}, the three-colored intensity diagram   
	clearly shows the relative spatial emission area and intensity traced by three isotopologue CO lines, respectively.

\subsection{Physical Properties of the CMC} \label{sec:physical properties}    
	 
\begin{figure}[htb]
		\centering
		\includegraphics[width=1\textwidth]{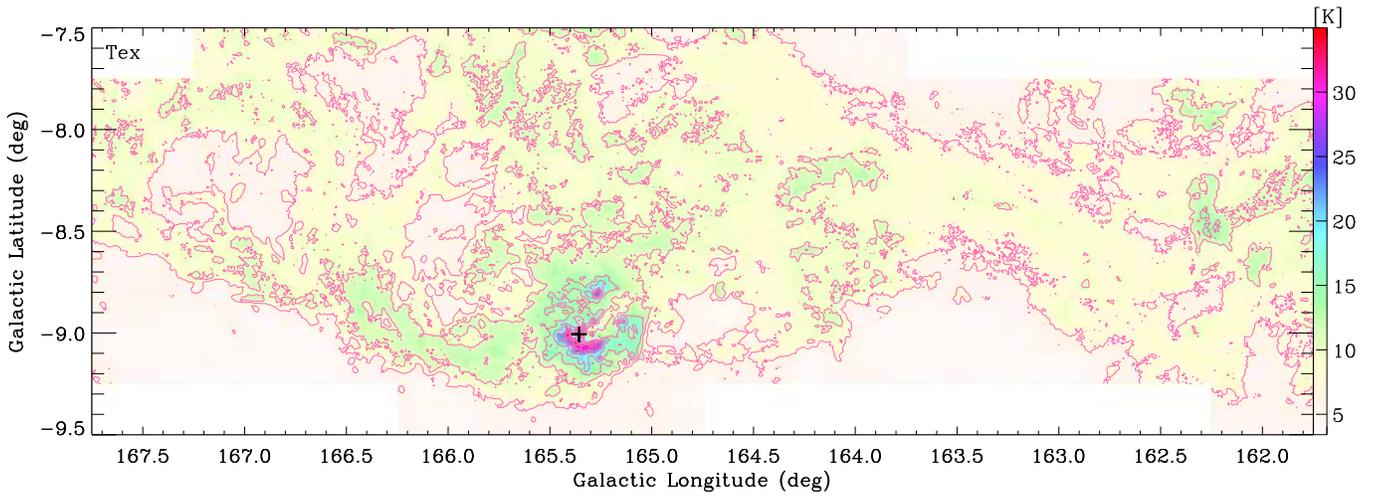}
		\caption{The excitation temperature map in the velocity interval of [-10, 5] km\,s$^{-1}$. 
		The contours are set from 3.4 K to 35.0 K by a step of 10\,\%. }
		 \label{fig-Tex}
	\end{figure}

	 We adopt 470 pc as the distance of the CMC from \cite{2019ApJ...879..125Z} who measured the distances of molecular clouds based on the dust-extinction
	 using the $Gaia$ DR2.
	 Assuming the \tw line is optically thick ($\tau \gg 1$), 
	 the basic physical parameters of the molecular cloud are estimated through the methods as follows.
	 Following \citet{1991ApJ...374..540G}, with the background temperature ${T}_{\rm bg}=2.73$ K, 
	 the excitation temperature ($T_{\rm ex}$) could be calculated from the peak temperature of the \tw emission ($T_{\rm mb, peak,^{12}CO}$):
	 \begin{equation}
	 T_{\rm ex}=5.532 \left({\rm log}(1+\frac{5.53}{T_{\rm mb,peak,^{12}CO}+0.819})\right)^{-1}~({\rm K}).
	 \label{eq:tex}
	 \end{equation}
	  As shown in Figure \ref{fig-Tex}, 
	  the excitation temperature ranges from 3.3 K to 34.9 K with a mean value of 7.8 K.
	  The Peak value (34.9 K) appears in the position of (l, b) = (165.34$\degr$, -9.08$\degr$), 
	  spatially coincides with the position of the LkH$\alpha$ 101 cluster. 
	  It is comparable with the dust temperature \cite[18-33 K; see][]{2017A&A...606A.100L}. 
	  The excitation temperature  around the subregion L1482 ($\sim$ 15 K) is higher compared to that around L1478 ($\sim$ 8 K).  
		 	
	  The H$_2$ column density can be estimated directly by integrating the main beam temperature ($T_{\rm mb}$)  of \tw over the velocity ($V$) 
	   using the CO-to-H$_2$ conversion factor $X =$ 1.8$\times$10$^{20}$~cm$^{-2}$~K$^{-1}$~km$^{-1}$~s from \cite{2001ApJ...547..792D}:
	  \begin{equation}
	  N_{\rm H_2} = X \int T_{\rm mb, ^{12}CO} dV.
	  \label{eq:N12}
	  \end{equation}
	  Under the assumption of local thermodynamic equilibrium (LTE) and 
	  optical thin of the $^{13}$CO and C$^{18}$O lines,
	  given the equal excitation temperature, the column densities traced by $^{13}$CO and C$^{18}$O  can be estimated from:
	  \begin{equation}
	  N_{\rm ^{13}CO} = 2.42 \times 10^{14} \cdot \frac{(1+0.88/T_{\rm ex})\int T_{\rm mb, ^{13}CO} dV}{1-{\rm exp}(-5.29/T_{\rm ex})}, \rm and
	  \label{N13}
	  \end{equation}  
	  \begin{equation}
	  N_{\rm C^{18}O} = 2.24 \times 10^{14} \cdot \frac{(1+0.88/T_{\rm ex})\int T_{\rm mb, C^{18}O} dV}{1-{\rm exp}(-5.27/T_{\rm ex})}.
	  \label{N18}
	  \end{equation}  
	  The H$_2$ column density could be then derived by multiplying \tht column density with the ratio of 
	  $N_{\rm H_2}/N_{\rm ^{13}CO}$ $\sim$ 7$\times$10$^{5}$ \citep{1982ApJ...262..590F} or multiplying \ei column density with the ratio of
	  $N_{\rm H_2}/N_{\rm C^{18}O}$ $\sim$ 7$\times$10$^{6}$ \citep{1995A&A...294..835C}.
	  The pixels at least 3 channels above 3\,$\sigma$ are kept in this calculation.
	  The column density derived from \tht span a range of 0.09$-$33.8 
	  $\times 10^{21} \mathrm{~cm}^{-2}$, a considerably larger range than those measured by \tw  
	   (0.10$-$19.6 $\times 10^{21} \mathrm{~cm}^{-2}$) 
	   and \ei (1.14$-$26.7 $\times 10^{21} \mathrm{~cm}^{-2}$).
	  In the estimation process of mass and number density of the CMC, a mean molecular 
	  weight per H$_{2}$ molecule of 2.83 \citep{2008A&A...487..993K} is adopted. The mean values are listed in Table \ref{tb-cloud}. 
	  
	  %the mass derived from \tht is about 3 times less than that derived from $^{12}$CO. 
	  The emission areas and the masses estimated from $^{12}$CO, $^{13}$CO, and C$^{18}$O gradually 
	  decrease in quality by an order of magnitude. 
	  The low-density molecular gas traced by \tw accounts for most part of the total mass, while the denser gas traced 
	  by \tht and \ei is better to trace the filamentary skeletons (see below).  
	  The use of a uniform abundance ratio %($N_{\rm H_2}/N_{\rm ^{13}CO}$ $\sim$ 7$\times$10$^{5}$)
	  will lead to a deviation in the estimation of mass.	
	  Actually, the abundances may be quite different in different regions.
	  Some studies have shown that under the assumption of optical thin and without considering the change of abundance, 
	  the mass calculated from \tht may be underestimated by 2-3 times \citep{2009ApJ...699.1092H,2008ApJ...680..428G}.
	   	 	
		\begin{deluxetable}{lccccc}
		\tabletypesize{\small}
		\setlength{\tabcolsep}{0.25in}
		\tablecaption{Properties of the CO emission in the CMC \label{tb-cloud}}
		\tablewidth{0pt} 
		%\tablenum{5}
		\tablehead{ \colhead{Emission} & \colhead{Velocity interval} & \colhead{Area} & \colhead{$N\rm(H_{2})$\tablenotemark{a}} & \colhead{$n\rm(H_{2})$\tablenotemark{a}} 
		& \colhead{$M\rm(H_{2})$}\\
		\colhead{}&\colhead{(km\,s$^{-1}$)} &\colhead{(deg$^{2}$)} & \colhead{($10^{21}$cm$^{-2}$)} & \colhead{(cm$^{-3}$)} & \colhead{($10^{4}$M$_{\sun}$)} 
		} 
		\colnumbers
		\startdata 
		$^{12}$CO(J=1-0)  & [-10, 5]  & 8.76     & 2.83 $\pm$ 0.19    & 67      & 2.59 $\pm$ 0.23     \\
		$^{13}$CO(J=1-0)  & [-10, 5]  & 4.85     & 2.14 $\pm$ 0.18    & 68      & 0.85 $\pm$ 0.33     \\
		C$^{18}$O(J=1-0)  & [-10, 5]  &  0.15    & 5.10 $\pm$ 1.84    & 911    & 0.09 $\pm$ 0.03     \\
		\enddata
		\tablecomments{a,  mean value with weight of integrated intensity.}
	\end{deluxetable}

\subsection{Filamentary Structure} \label{Filamentary Structure}	

\subsubsection{Major Filament} \label{Major Filament}
	
	As shown in Figure \ref{fig-lv}, the \tht emission distributes continuously in the velocity interval of [-5, 3] km\,s$^{-1}$. 
	We calculate the column density in this velocity interval and apply for the DisPerSE algorithm \citep{2011MNRAS.414..350S} 
	to extract the skeleton structures. 
	In the process of the identification, the persistence threshold  is set to be $5.8\times 10^{21}$ cm$^{-2}$ ($\sim$ 50\,$\sigma$). 
	The parameter -trimBelow used to filter noise is set to be $3.5\times 10^{20}$ cm$^{-2}$ ($\sim$ 3\,$\sigma$).  
	This setting would not only guarantee a clear skeleton construction with a highly contrast with the surrounding, 
	but also retain the information of persistence in morphology as far as possible.
	As shown in Figure \ref{fig-m0},
	%where the black line indicates the skeleton.  %the gray lines indicate the branches identified by DisPerSE algorithm.
	the skeleton construction extends from 162.2$\degr$ to 166.6$\degr$ and shows an integral symbol-like shape. 
	Adopting the distance of 470\,pc, the length of the filament is estimated to be $\sim$ 72 pc. 
	Hereafter, we name this filament as California Molecular Filament (CMF).
	
	We construct the mean column density profile $N_{p}(r)$ of the CMF from radial cuts using a similar procedure as described in 
	\cite{2011A&A...529L...6A} and \cite{2013A&A...550A..38P}. 
	Applying for the Plummer-like function $N_{p}(r)= N_{\mathrm{H}_{2}}^{0} /\left[1+\left(r / R_{\mathrm{flat}}\right)^{2}\right]^{(p-1)/2}$ \citep{2016A&A...590A.110C},
	where $N_{\mathrm{H}_{2}}^{0}$ is the amplification,
	$R_{\rm flat}$ is the inner flattening radius, and $p$ is the power index at large radius $\geqslant$ $R_{\rm flat}$,  
	to the radial density profile, the $R_{\rm flat}$ is estimated to be $\sim$ 0.30 $\pm$ 0.01 pc and $p$ is $\sim$ 1.62 $\pm$ 0.01. 
	The full width at half maxima (FWHM) of the Plummer-like profile ($2\,R_{\mathrm{flat}} \sqrt{2^{2 /(p-1)}-1}$) is $\sim$ 1.73 pc,
	which is equivalent to $\sim$ 3 $\times$ (2\,$R_{\rm flat}$).
	The  Gaussian FWHM of the averaged \tht spectrum ($\Delta V$) is $\sim$ 1.95 $\pm$ 0.22 km\,s$^{-1}$. % (see Figure \ref{fig-fil}). 
	The observed velocity dispersion $\sigma_{\rm obs}=\Delta V /\sqrt{8\ln2}$ is thus to be % ($\Delta V$ is FWHM) derived
	$\sim$ 0.83 $\pm$ 0.10 km\,s$^{-1}$.
	
\begin{figure}[htb]
		\centering
		\includegraphics[width=0.5\textwidth]{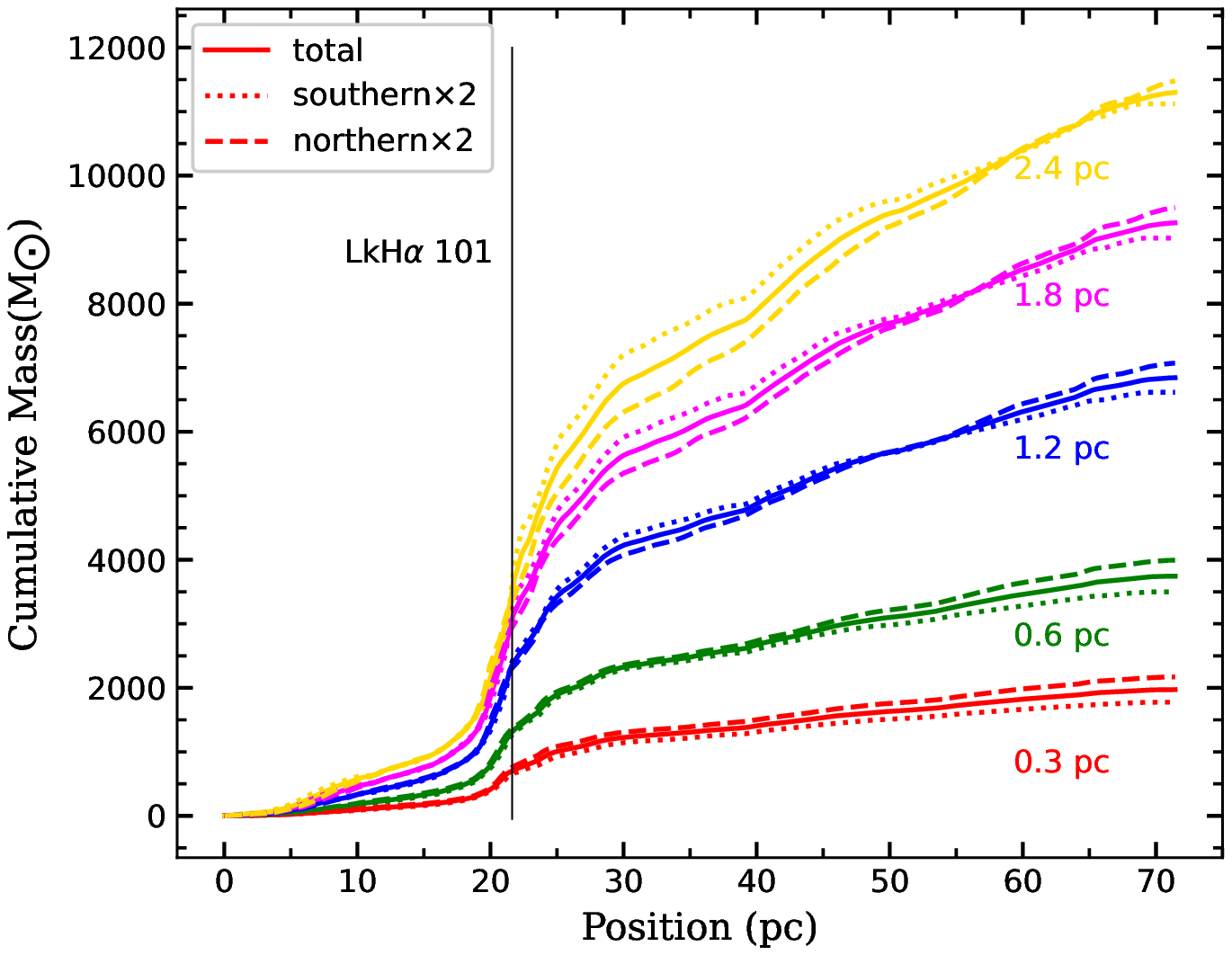}%\textit{(a)}
		\includegraphics[width=0.5\textwidth]{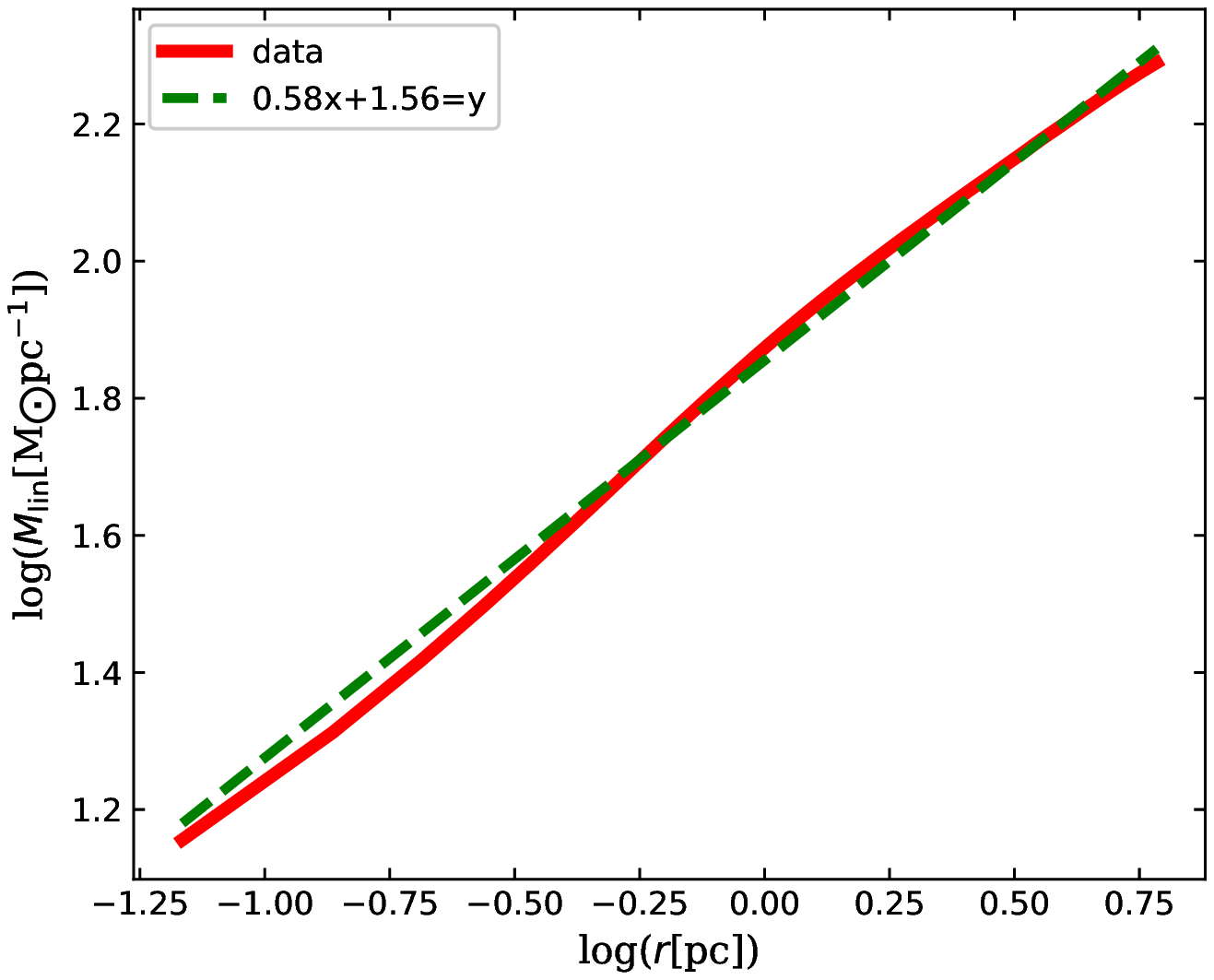}\\%\textit{(a)}
		\caption{Left panel: The cumulative mass distribution along the CMF at different projected radii from the ridgeline, 
		which is indicated by different colors.  
		The profiles of southern and northern CMF multiplied by 2 are indicated by dotted and dashed lines, respectively.
		The LkH$\alpha$101 is indicated by vertical  line. 
	        Right panel: The enclosed line-mass profile (red curves) as a function of the projected radius and the fitting line (green dashed line) 
	        in logarithmic spaces. 
	       }
		\label{fig-cum}
	\end{figure}

	Similar to the Figure 6 in \cite{2021ApJ...908...86A}, we show the cumulative mass profile along 
	the CMF at the projected radii of  $R_{\rm flat}$, 2$R_{\rm flat}$, 4\,$R_{\rm flat}$, 6\,$R_{\rm flat}$, and 8\,$R_{\rm flat}$ 
	on the north and south sides of the ridgeline respectively in Figure \ref{fig-cum}.  
	The profiles show steep rises near the LkH$\alpha$ 101 cluster, indicating that the mass distribution along the CMF is inhomogeneous. 
	The mass distributions are broadly symmetrical within 2$R_{\rm flat}$, being consistent with the result of \cite{2021ApJ...908...86A} for the 
	subregion L1482. 
	As shown in the right panel of Figure \ref{fig-cum},  the enclosed average 
	cumulative line-mass profile along the minor axis is extracted to check the dynamic state of the CMF as a whole.  
	By applying for a power-law fit $M_{\rm line}(r)=f\times\left(\frac{r}{\mathrm{pc}}\right)^{\gamma}$ to the profile, 
	a power index $\gamma$ of 0.58 is derived , which is comparable with that estimated from the subregion L1482 (0.62).
	As shown in  Figure \ref{fig-mline},  the $M_{\rm line}$ profile along the CMF at the radii of 0.6 pc and 2.4 pc are showed respectively.

	\begin{deluxetable}{cccccccc}
		\tabletypesize{\small} 
%		\tablecolumns{5} 
		\setlength{\tabcolsep}{0.15in}
		\tablecaption{Properties of the CMF \label{tb-filaments}}
		\tablewidth{0pt}
		\tablehead{ \colhead{Emission} & \colhead{Length} & \colhead{Width\tablenotemark{a}} & \colhead{Mass} & 
			\colhead{$ T_\mathrm{ex} $} & \colhead{$\Delta V$}&
			\colhead{$ N_\mathrm{H_2} $} & \colhead{$M_\mathrm{line}$} \\
			& \colhead{(pc)} & \colhead{(pc)} & \colhead{(\msun)} & \colhead{(K)} 
			& \colhead{(km\,s$^{-1}$)}  & \colhead{($ 10^{21} $ cm$^{-2}$)}  & \colhead{(\msun\,pc$^{-1}$)}}
		\colnumbers
		\startdata
		\tht &  71.8     & 0.60  $\pm$ 0.02   & 1793 $\pm$ 144          &  10.4 $\pm$ 3.1  & 2.0 $\pm$ 0.2  & 3.64 $\pm$ 0.18   & 26\\
		\enddata 
		\tablenotetext{a}{Derived from Plummer-like function fitting.}
		%\tablenotetext{b}{FWHM.}
	\end{deluxetable} 	
		
	As shown in Figure \ref{fig-channel_L}, 
	the filamentary structure appears in the velocity range from -2.75 to 1.25 km\,s$^{-1}$. 
	The molecular gas to the north side of the CMF mainly concentrated in the velocity of $v \textless$ -1.25 km\,s$^{-1}$,
	while the gas gather along the CMF has $v_{\rm LSR}$ range from -1.25 to 0.25 km\,s$^{-1}$.
	The western CMF ($l \textless$ 164.5$\degr$) and eastern CMF ($l \textgreater$ 164.5$\degr$) seem to have different systemic velocities, 
	$\sim$ [-1.75, -0.75] km\,s$^{-1}$ for the eastern and $\sim$ [-1.25, -0.25] km\,s$^{-1}$ for the western, respectively.

	In order to obtain a detailed understanding of the filament kinematics, we show the velocity distribution (m1) map
	of \tht in  [-2, 0] km\,s$^{-1}$ in Figure \ref{fig-m1_L}.  
	The gas surrounding the integral-shaped filament is redshifted in the south side and blueshifted in the north side. 
	As shown in Figure \ref{fig-pv_ver},  ten PV slices are extracted across the ridgeline of the CMF. 
	The velocity structure of  \tht  across the major axis (horizontal black lines) is characterized by a velocity difference
	going from $v_{\rm LSR}$ $\sim$ -2 km\,s$^{-1}$ to $v_{\rm LSR}$ $\sim$ 0 km\,s$^{-1}$ along a length of about 1 pc. 
	The fitted velocity gradients range from 0.66 to 1.14 km\,s$^{-1}$\,pc$^{-1}$, with a mean value 
	of $\sim$ 0.82 km\,s$^{-1}$\,pc$^{-1}$. 
	  	
\subsubsection{Substructures} \label{Sub-structures}	
		
	Compared with \tw and $^{13}$CO, the \ei emission traces the innermost structures of the molecular cloud. 
	 We try to combine the DisPerSE algorithm with the channel column density maps to resolve \ei structures. 
	 The \ei emission is detected in the region with bright \tht emission, and mainly appears
	 at $v_{\rm LSR} \sim$ -1 km\,s$^{-1}$
	 (see Figure \ref{fig-lv} and Figure \ref{fig-m0}). 
	 The velocity resolution in the observations is $\sim$ 0.17 km\,s$^{-1}$, which
	 is comparable to the isothermal sound speed in the cloud. 
	 Based on the assumption that a filament should be detected in more than three consecutive velocity channels, 
	 we estimate the channel column density every three velocity channels (0.5 km\,s$^{-1}$) with pixels above 3\,$\sigma$ to extract the 
	 the filamentary structures. 
	 Then we merge these identified filamentary structures into a single intensity image (see Figure \ref{fig-fiber-pv}).
	 The identified skeletons in the six velocity intervals are marked out by different colors. 
	
	 As shown in Figure \ref{fig-fiber-pv}, we zoom in the Eastern 
	 ($164.7^{\circ} \leqslant l \leqslant 166^{\circ},-9.3^{\circ} \leqslant b \leqslant-8.2^{\circ}$) and Western 
	 ($162.2^{\circ} \leqslant l \leqslant 164.5^{\circ},-9.1^{\circ} \leqslant b \leqslant-7.9^{\circ}$) regions (hereafter "E" and "W") and
	overlay the YSOs from \cite{2017A&A...606A.100L} on them. 
	By the position of the cluster LkH$\alpha$ 101, E could be further divided into SE ($l \sim$ [165.5\degr, 166.0\degr]), 
	NE1 ($l \sim$ [165.1\degr, 165.5\degr]), and NE2 ($l \sim$ [164.8\degr, 165.1\degr]),  respectively.
	W is divided into NW ($l \sim$ [163.8\degr, 164.5\degr]), SW1 ($l \sim$ [162.6\degr, 163.2\degr]), SW2 ($l \sim$ [163.2\degr, 163.8\degr]), and 
	 X-structure ($l \sim$ [162.2\degr, 162.6\degr]),  respectively.
	Along the directions of the arrowed lines,
	 two PV plots are extracted from the \ei  m1 map (with pixels above 3\,$\sigma$).   
	Combing the distributions of the substructures in position-position (PP) and PV spaces, we could have a better understanding 
	of their dynamics. The \ei emission in E appears to be concentrated at -1 km\,s$^{-1}$ (green).
	The spiny structures constructed by a few pixels (maroon or cyan in SE,  red or blue in NE) frequently seen
	 along the PV plot in our study.
        In NE1,  the emission of the substructures is blueshifted in the north side and redshifted in the south side.
	It can be seen that there are at least three substructures with $v_{\rm LSR}$ at -2, -1.5, and -1 km\,s$^{-1}$ parallel 
	distributing within the CMF in "zig-zag" shape.
	In this region, there is a distinct velocity gradient from blueshifted to redshifted  towards the cluster LkH$\alpha$ 101. 
	In NE2, more substructures are seen to overlap and merge to form hub-like features, and the YSOs in this region are more concentrated than  NE1.
	At the western end of SE, where YSOs are the most concentrated, the substructures also present hub-like morphological features. 
	The emission in W appears to be concentrated at -1.5 km\,s$^{-1}$ (cyan) and -2 km\,s$^{-1}$ (blue).
	Compared with E, fewer spiny structures appear in this region. %The spiny structures are rare than that in E.
        In SW1, the substructures with $v_{\rm LSR}$ at $\sim$ -1.5, -2 km\,s$^{-1}$ are overlaid and their velocities are coherent in a wave form.
	In SW2,  the substructures with $v_{\rm LSR}$ at $\sim$ -1.5 km\,s$^{-1}$ and $v_{\rm LSR}$ at $\sim$ -2 km\,s$^{-1}$ 
	are spatially separated and their velocity distributions are more diffuse than that in SW1. 
	In NW, the substructures appear to fragment independently of each other.
	For the X structure, the substructures with $v_{\rm LSR}$ at $\sim$ -2.0, -1.5 kms$^{-1}$ and 
	$v_{\rm LSR}$ at $\sim$ -0.5, 0.0 km\,s$^{-1}$ are located in north and south, respectively.	
	
\subsection{Identification of Dense Cores} \label{sec: Prestellar cores}		

	The theoretical models of infinite self-gravitating fluid cylinder suggest that when the line mass ($M_{\rm line}$) of an 
	isothermal filament exceeds the critical value for equilibrium, 
	the entire filament will collapse towards the axis and then fragment into clumps or 
	cores \citep{1953ApJ...118..116C,1964ApJ...140.1056O,2010ApJ...719L.185J}.
	By examining dense cores along filaments, 
	one could understand the evolutionary phases of the filament and star formation therein.	
	Thus, we use the Gaussclumps procedure in GILDAS \citep{1990ApJ...356..513S}  to 
	search for dense cores from \ei ([-4, 1] km\,s$^{-1}$). The threshold is set to be 0.87\,K (3\,$\sigma$) to 
	ensure the completeness of the sample. Other parameters adopt the recommended values from \cite{1998A&A...329..249K}.
	After removing the cores that with short axes 
	less than the spatial scale (0.12\,pc) and that are located at the edges of the observed region,
	we extract 225 \ei cores from the original sample of 5000 cores.
	Assuming that each core has a Gaussian-shaped density distribution both in spatial and velocity coordinates, 
	the Gaussclumps algorithm derive the parameters of the Longitude and Latitude of Galactic coordinate system,   
	angular sizes of the major axis and minor axis, velocity, peak temperature and line width (FWHM).
	By the methods as described in \cite{2016A&A...588A.104G}, the physical properties of the dense cores including the deconvolved 
	radius ($R\rm_{eff}$), excitation temperature ($T\rm_{ex}$), column density ($N\rm_{H_{2}}$),
	number density ($n\rm_{H_{2}}$), molecular cloud mass under local equilibrium ($M\rm_{LTE}$), virial mass ($M\rm_{vir}$)
	and virial parameter ($\alpha\rm_{vir}$ = $M\rm_{vir}$/$M\rm_{LTE}$) are derived and 
	tabulated in Table \ref{tb-clumps}. 
	%The virial mass was calaculated by $M_{vir} = 209 Reff \Delta v^2$. 
	
	The effective radius after deconvolution of the dense cores range from 0.02 pc to 0.28 pc with a mean value of 0.12 pc. 
	The excitation temperatures span a range of 6.1$-$33.2 K with a mean value of 9.8 K.
	The number density ranges from  0.22$\times 10^{4}$ cm$^{-3}$ to 29.4$\times 10^{4}$ cm$^{-3}$ with a mean value 
	of 1.47$\times 10^{4}$ cm$^{-3}$.
	The mass under local equilibrium ranges from 0.65 M$_{\sun}$ to 53 M$_{\sun}$ with a mean value of 4.5\,\msun.
	The virial parameters span a range of 0.30$-$2.49 with a mean value of 1.15.
	The ratios of cores with $\alpha_{\mathrm{vir}} \textless 1$ and $\alpha_{\mathrm{vir}} \textless 2$ are 
	approximately $41\,\% \,(93 / 225)$ and $96\,\% \,(215 / 225)$, respectively. Most of the dense cores are in 
	gravitationally bound states. 
	We remove 39 cores with overlapping the YSOs as well as 5 cores with associated $Herschel$ 70 $\mu$m 
	point-sources \citep{2013ApJ...767...36S}.
	Finally, we get 181 starless dense cores from the \ei emission. 
	For comparison, Table \ref{tb-comparecores} list the median values of the calculated physical parameters with 
	the 95 \% percentile and 5 \% percentile for cores with YSOs and starless cores, respectively.
	The group of starless cores have a relatively lower effective radius, excitation temperature, number density, 
	 mass and  a slightly larger $\alpha\rm_{vir}$ comparing to the dense cores that overlap with YSOs. 	
			
\subsection{Identification of Outflow Candidates} \label{sec: outflow}		
		 
	 Molecular outflow is a direct probe of star-forming activities \citep{2007prpl.conf..245A}.
	 The low-J pure rotational transitions of CO have been the most used tracers of molecular outflows 
	 as it is abundant [$N$(CO) $\sim$ 10$^{-4}$ $N$(H$_{2}$)] and could be easily populated by collisions with 
	 H$_{2}$ and He at the typical densities and temperatures of molecular clouds \citep{2016ARA&A..54..491B}. 
	 In order to investigate star formation along the CMF, we try to search for outflows from the \tw and \tht emission by checking 
	 protruding structures in the PV diagrams and mapping the spatial distribution of integrated line wings. 
	 %We make a PV slice along the major axis of the CMF from east to west (see Figure \ref{fig-pvshow}). 
	 As shown in Figure \ref{fig-pvshow}, 
	 there are about 15 positions (black boxes) exhibiting protruding structures which can be regarded 
	 as candidate regions for outflows.  
	 We extract the spectral lines with four pixels around the positions 
	 and estimate their center velocities $v_{\rm LSR,cen}$ as a reference of the $v\rm_{center}$ of the outflow candidates 
	 by applying for the Gaussian fitting to \tht lines. 
	 Then we gradually increase velocity intervals between the $v\rm_{center}$ and the inner edges of the red or blue lobes 
	 until the extrapolated integrated intensity maps of the lobes could be distinguished from the background intensity maps. 
	 We map the integrated red or blue lobes in the 15 regions and show them in Figure \ref{fig-outflow}. 
	 The average intensity of the noise level is estimated by 
	 $\sigma$(blue/red) = $\sigma \times \sqrt{V_{range}(\rm blue/red)\times 0.17}$,
	 where $\sigma$ is the noise level and 0.17 is the velocity channel in unit of km\,s$^{-1}$.
	 \cite{2017A&A...606A.100L} combined two extensive catalogs of YSOs in the CMC \citep{2013ApJ...764..133H,2014ApJ...786...37B}
	 to provide a census of YSOs which are classified as P: protostar (Class 0/I), D: disk (Class II) and S: star (Class III). 
	 To investigate the potential origin of these outflow candidates, we overlay these YSOs on the zoomed-in maps. 
	 The Galactic coordinate, the $v\rm_{center}$, the velocity intervals of the blueshifted and redshifted lobes, 
	 the times of the integrated noise level, the center locations of the outflow candidates, the approximate boundary 
	 locations of the red (blue) lobes, and 
	 the associated YSOs in the 15 regions are listed in Table \ref{tb-outflow}. 
	 As shown in Figure \ref{fig-outflow},  about 20 outflow candidates with 31 individual lobes are identified.
	 Among them, except the candidates numbered 6 and 17, the others are seen to match well with YSOs or dense cores. 
	 As shown in Figure \ref{fig-cores_2}, the sketch map of the distribution of these candidates shows that  
	 the angle between the outflow candidate and the major axis of the CMF seems to have a random distribution.
			
\section{Discussions} 	
\subsection{Large-Scale Major Filament in the CMC} \label{sec: large scale filament}	
	
\begin{figure}[htb]
		\centering
		\includegraphics[width=1.0\textwidth]{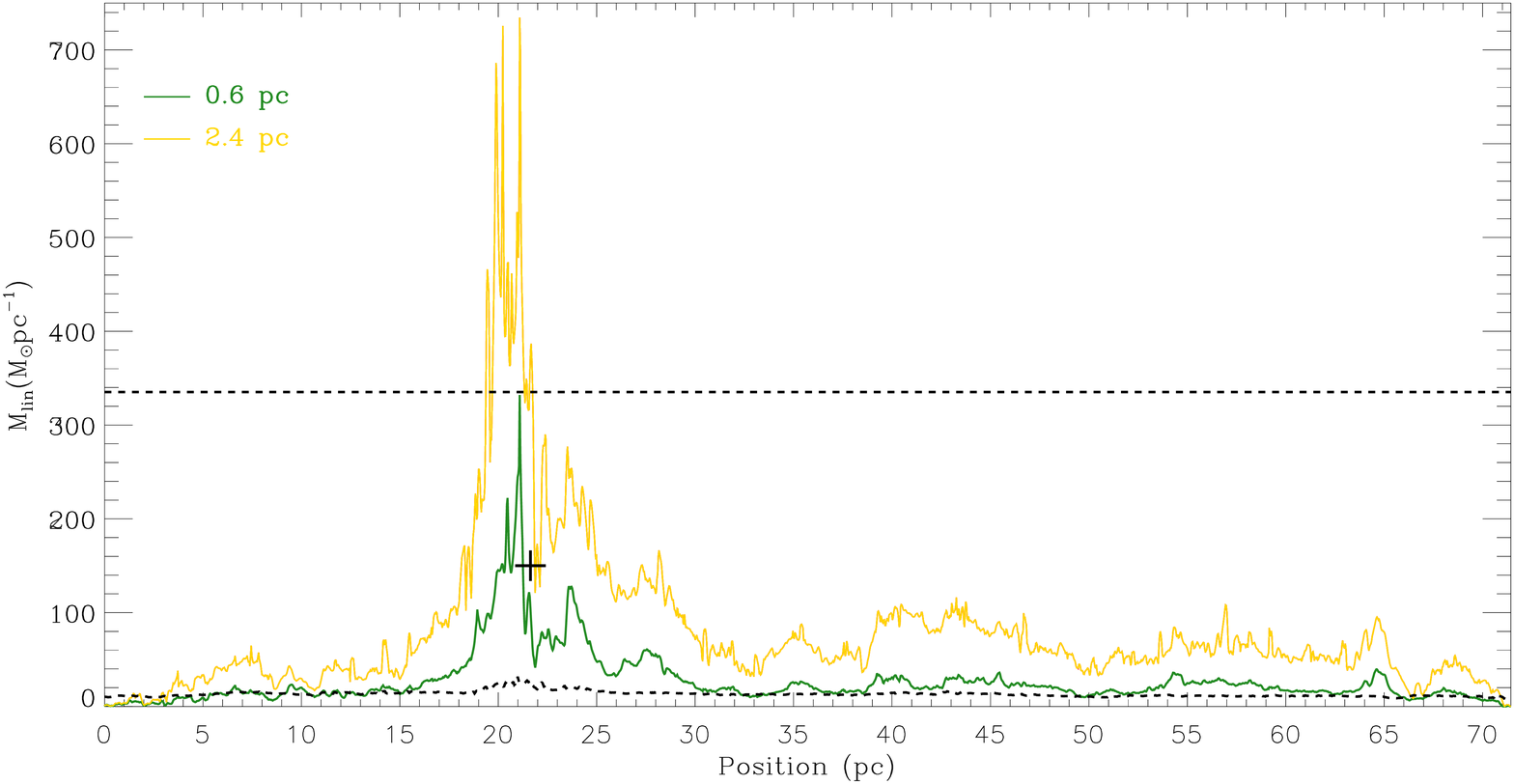}%\textit{(a)}
		\caption{The $M_{\rm line}$ profiles at the projected radii of 0.6\,pc (green curve) and 2.4\,pc (gold curve) along the CMF.  
		The lower and upper dashed lines show the thermal supported $M_{\rm line, crit}$ and turbulent supported $M_{\rm line, crit}$, respectively.		
		Most parts of the CMF have $M_{\rm line}$  greater than the thermal supported $M_{\rm line, crit}$ (see the green curve), 
		the regions around the LkH$\alpha$ 101 have $M_{\rm line}$ greater than turbulent supported $M_{\rm line, crit}$ (see the gold curve), 
		which suggests fragmentation processes taking place along the CMF due to gravitational instability.
	        }
		\label{fig-mline}
	\end{figure}

\begin{figure}[htb]
		\centering
		\includegraphics[width=1.0\textwidth]{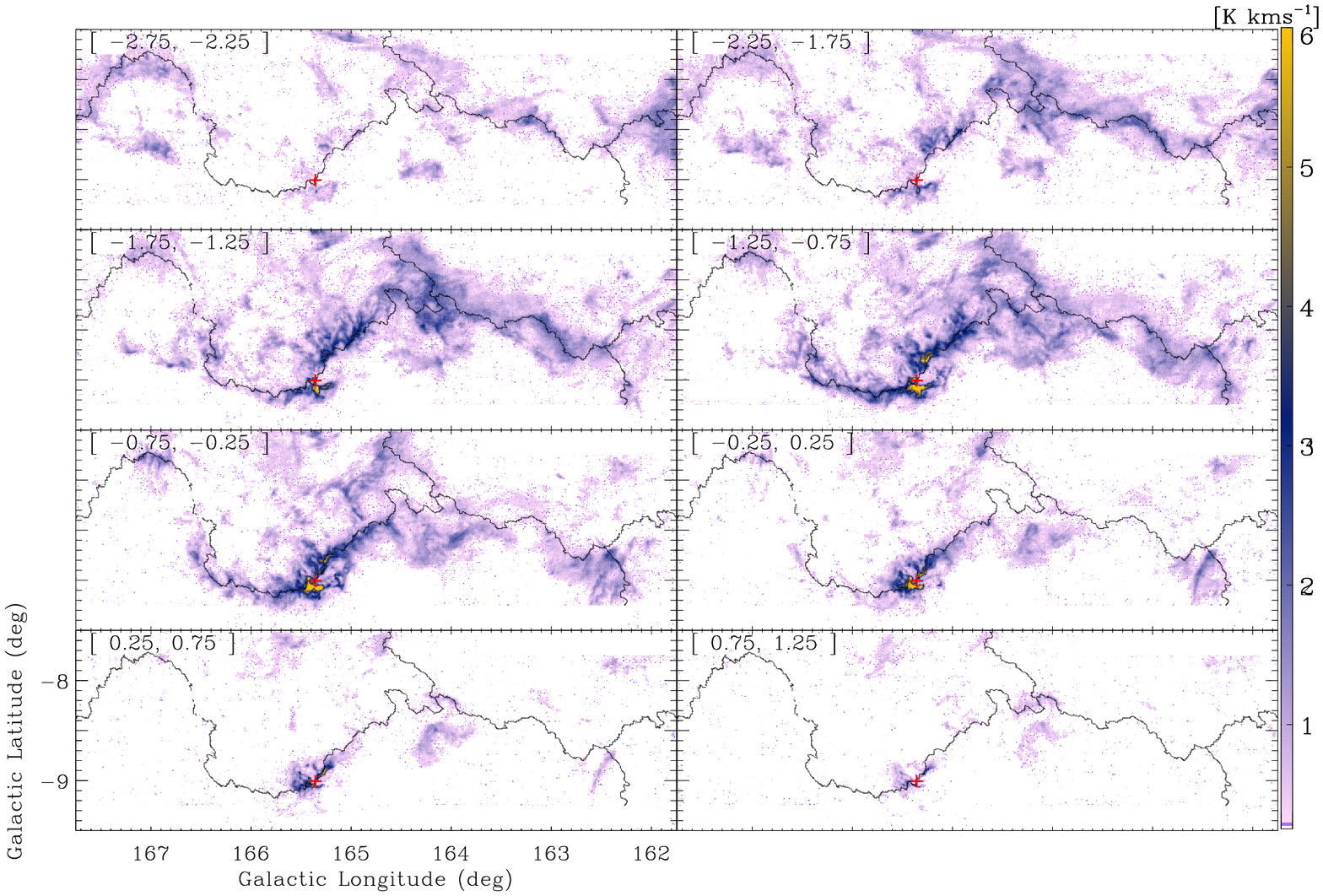}\\%\textit{(a)}
		\caption{The \tht  velocity channel maps from -2.75 to 1.25 km\,s$^{-1}$  by a step of 0.5 km\,s$^{-1}$.
		The velocity intervals are marked in the top-left corner of each panel.
		The colorbar is in the unit of K km\,s$^{-1}$. The overlaid black lines indicate the skeleton extracted by DisPerSE.
		The channel maps show the kinematic of the CMC in detail.
		}
		\label{fig-channel_L}
	\end{figure}

\begin{figure}[htb]
		\centering
		\includegraphics[width=1.0\textwidth]{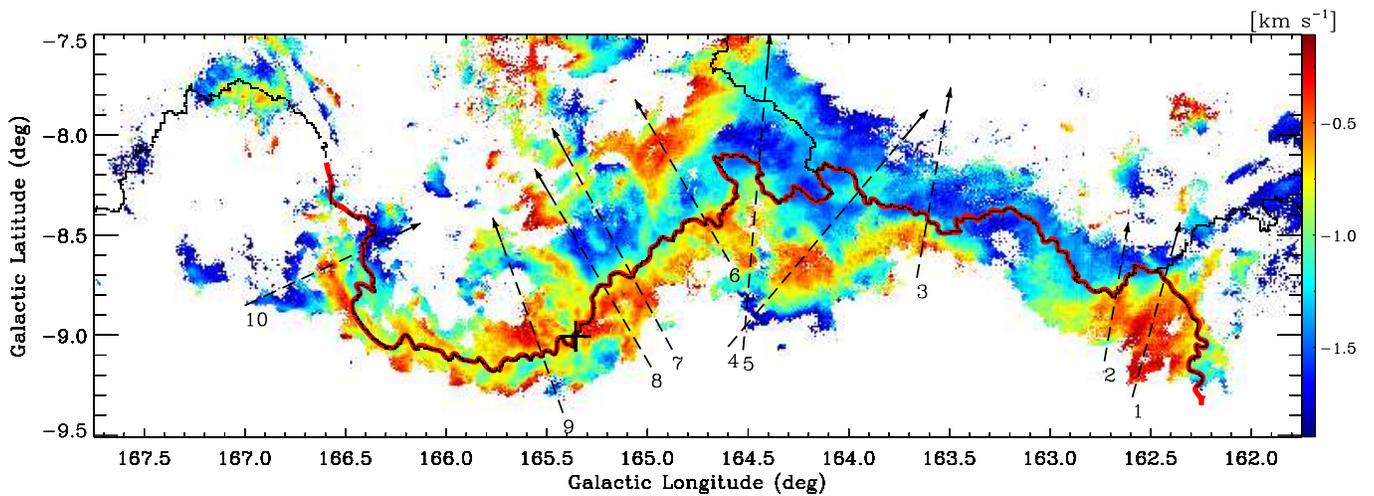}%\textit{(a)}
		\caption{The m1 maps of \tht in the velocity interval of [-2, 0] km\,s$^{-1}$. 
		The red curve indicates the ridgeline  of the CMF. 
		 The black curves are same as Figure \ref{fig-channel_L}.
		The black dashed lines perpendicular to the ridgeline are the paths of the nine PV slices.}
		\label{fig-m1_L}
	\end{figure}

\begin{figure}[htb]
		\centering
		\includegraphics[width=1.0\textwidth]{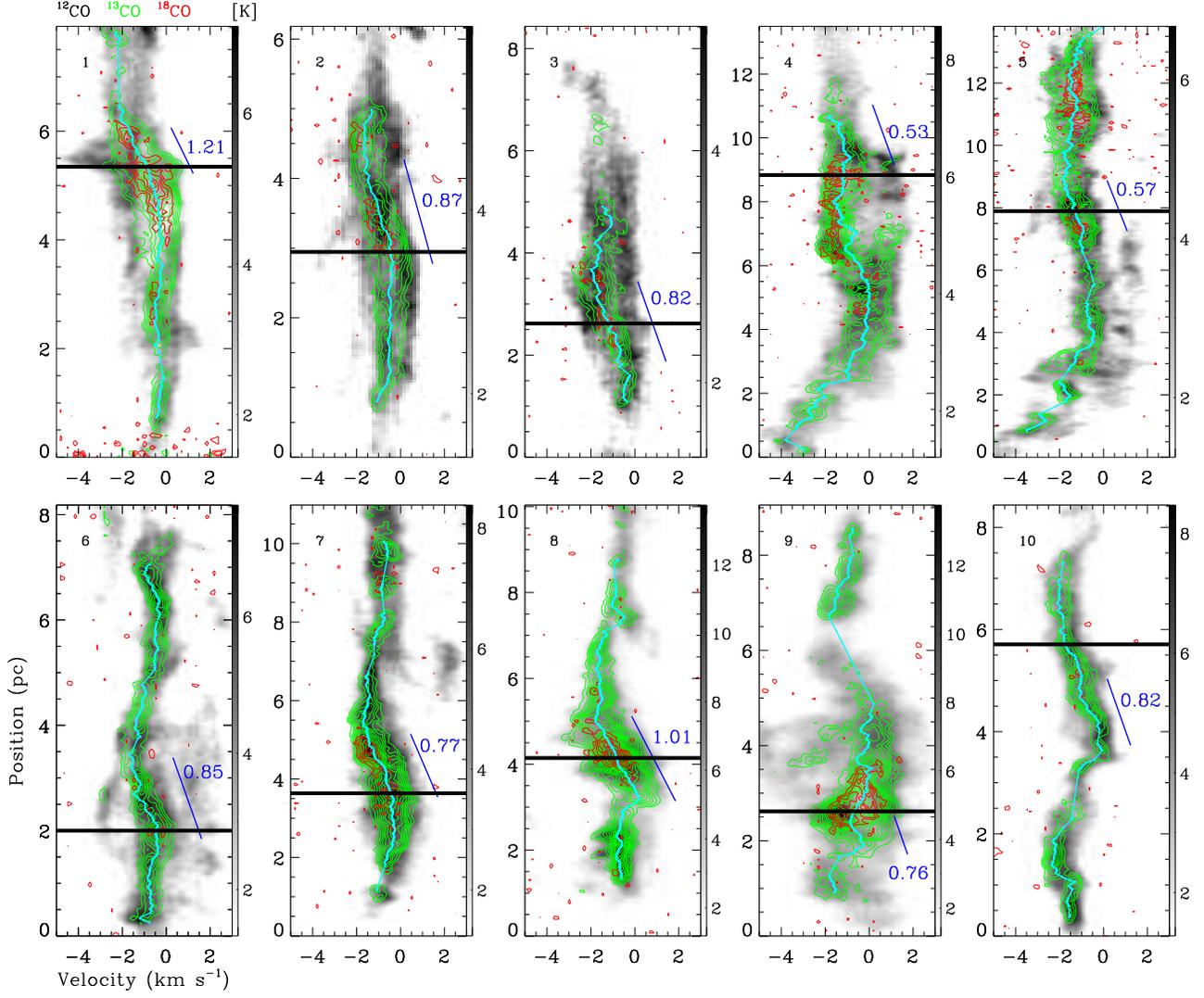}%\textit{(a)}
		\caption{Ten PV plots with widths of 2 pixels along the paths indicated in Figure \ref{fig-m1_L}. 
		The grayscale background,  green contours, and red contours represent  the
	     $^{12}$CO, $^{13}$CO, and C$^{18}$O emission, respectively.
	     The green contours in each map are set from 3\,$\sigma$ ($\sim$ 3 $\times$ 0.28 K) to peak value 
	     in an average step of 10\,\%.   
	    The red contours in each map are set from 1.5\,$\sigma$ to peak value in an average step of 20\,\%.  
	     The colorbar is in the unit of K.
	     The cyan curves indicate the \tht m1 profiles.
	     The horizontal black lines represent the positions that the CMF come cross each slice.
	     The blue lines are the linear fitted results from the segments of m1 profiles.
	     The corresponding fitted velocity gradients in the unit of km\,s$^{-1}$pc$^{-1}$ are marked besides.}
		\label{fig-pv_ver}
	\end{figure}

\begin{figure}[htb]
		\centering
		\includegraphics[width=1.0\textwidth]{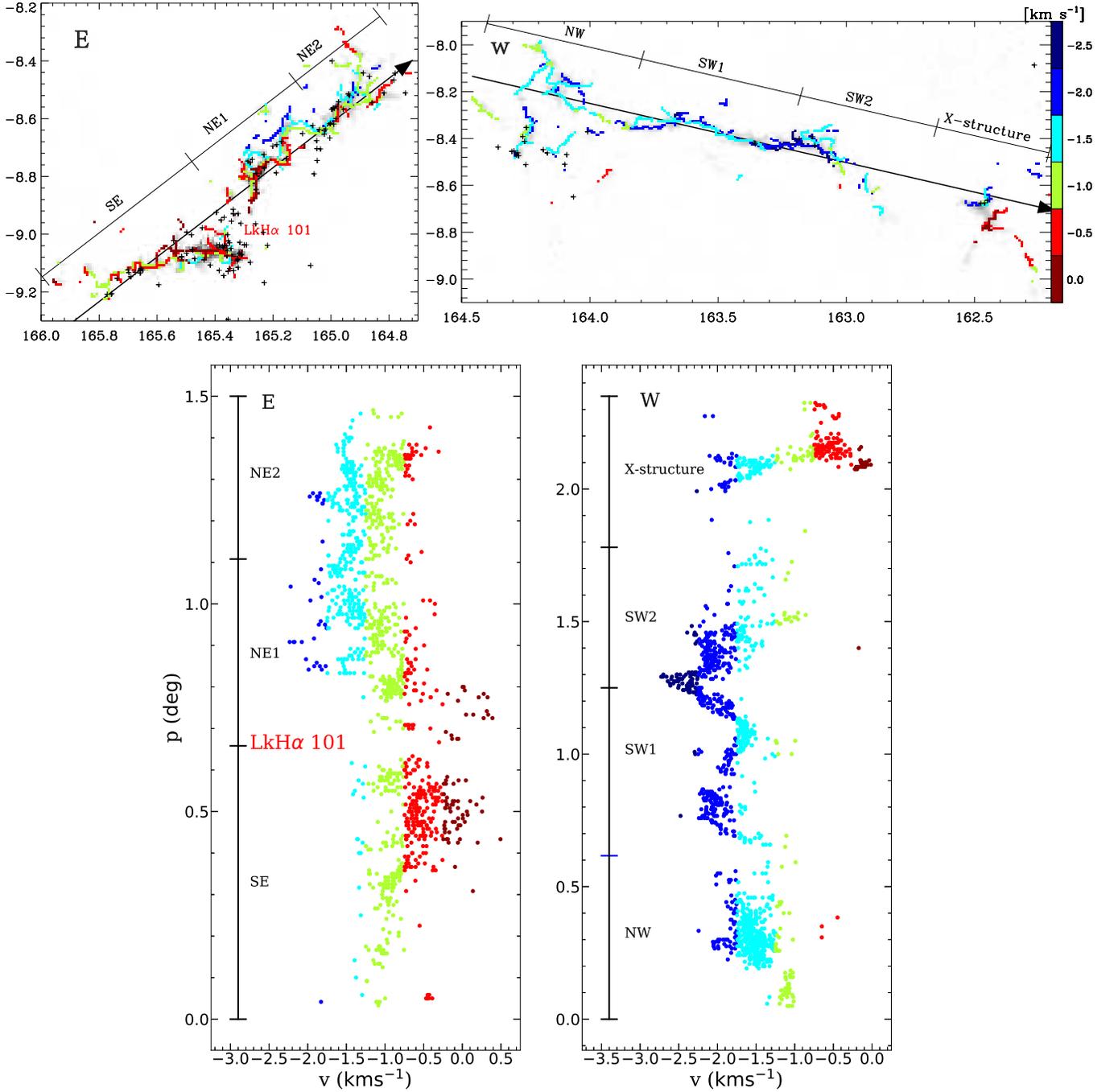}\\%\textit{(a)}
		\caption{Upper panel: The merged skeletons (marked by different colors) extracted from the \ei emission in region E and W.
		Each skeleton is extracted  from the channel $\mathrm{H}_{2}$ column density maps 
		( from -2.75 to 0.25 \,km\,s$^{-1}$ by a step of 0.5\,km\,s$^{-1}$ with pixels above 3$\sigma$ kept) by the DisPerSE algorithm.
		The plus symbols indicate YSOs. The black lines indicate the extents of the regions defined in section $3.3.2$.
		Bottom panel: The PV plots of m1 distribution (with pixels above 3 $\sigma$ kept) in region E and W.
		 The color setting and the extents in the black vertical lines are same as that of the upper panel.
		  }
		\label{fig-fiber-pv}
	\end{figure}

\begin{figure}[htb]
		\centering
		\includegraphics[width=1.0\textwidth]{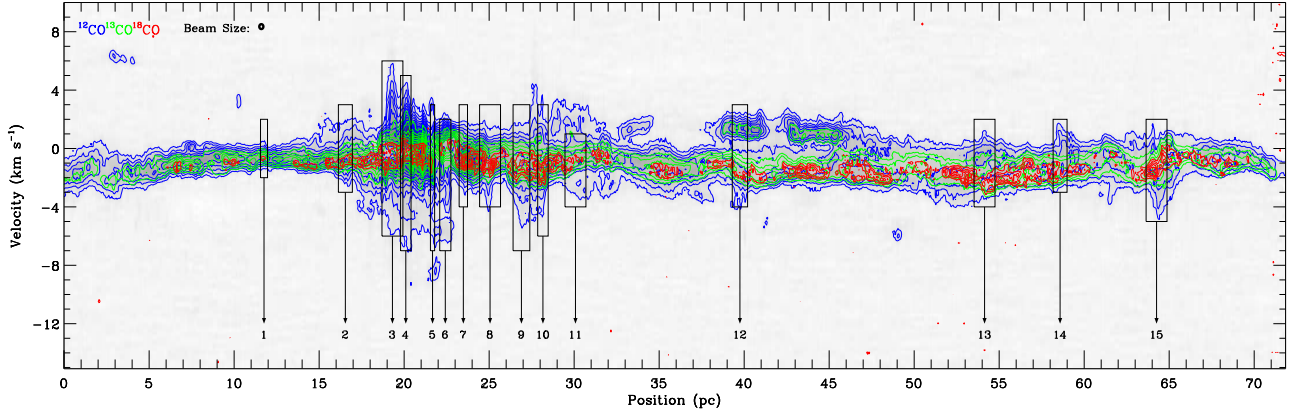}%\textit{(a)}
		\caption{ 
		The PV diagram along the major axis of the CMF in the direction from east to west. 
		 The blue and green contours represent the emission of \tw and \tht from 3\,$\sigma$ to peak 
		 value by a an average step of 3\,$\sigma$. 
		 The red contours represent the emission of \ei from 1.5\,$\sigma$ to peak 
		 value by a an average step of 1\,$\sigma$.
		 %1\,$\sigma$ is approximately equal to 0.44 K for \tw, 0.28 K for \tht, and 0.26 K for \ei, respectively.
		The black boxes mark out the positions of the protruding structures along the CMF.}
		\label{fig-pvshow}
	\end{figure}

\begin{figure}[ht]
		\centering
		\includegraphics[width=0.9\textwidth]{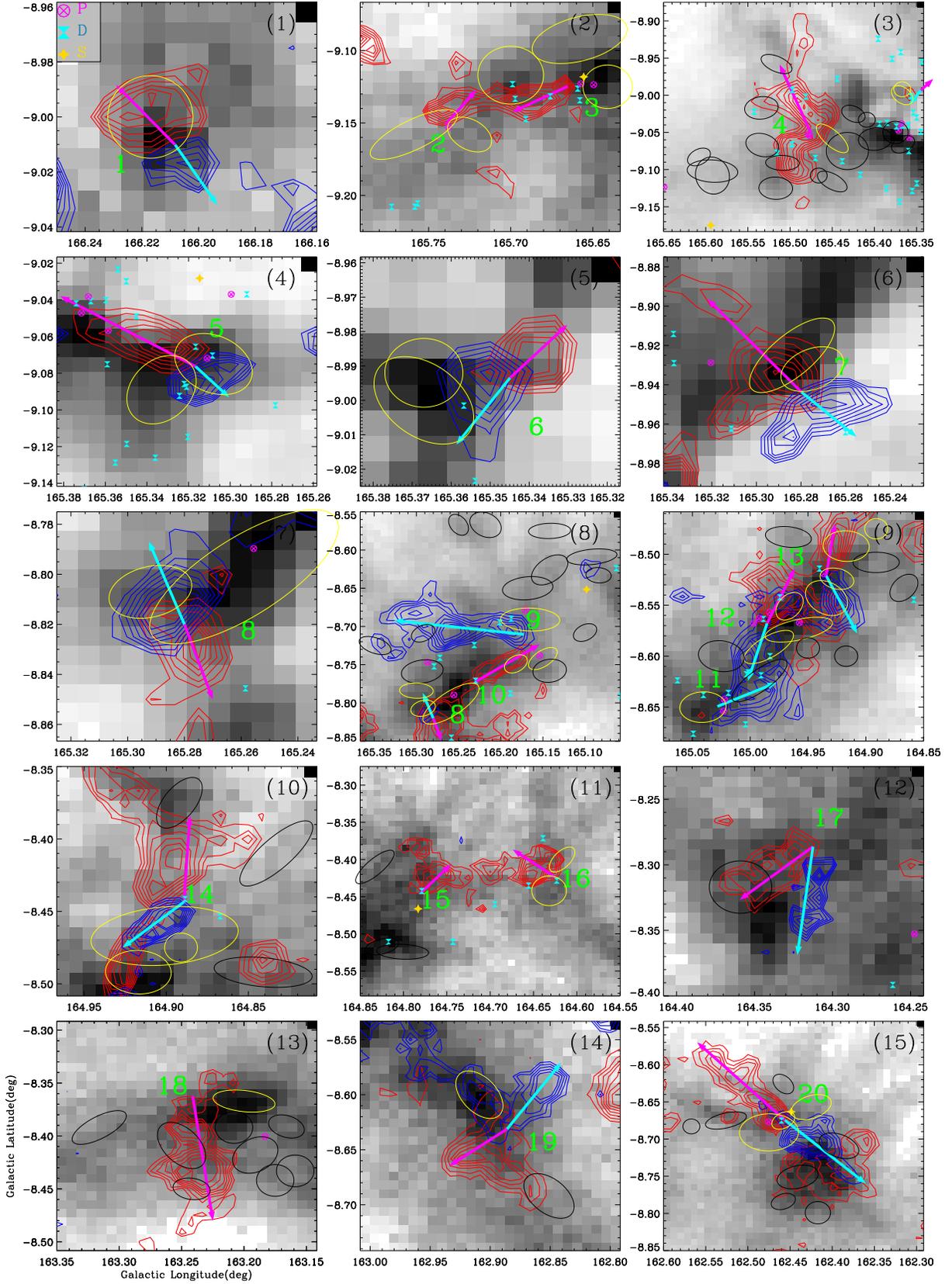}
		\caption{The 15 zoom-in regions with the red (blue) lobes as the overlaid red (blue) contours. 
		The ID of the outflow candidates are marked besides.
		The background is the \tht intensity map.
		The YSOs define P, D, S are also overlaid on each map.
		The ellipses are the \ei dense cores in which yellow represents the ones whose central locations are within 2.5$\arcmin$ from the 
		driving positions of the outflow candidates and black represents the opposite.}
		\label{fig-outflow}
	\end{figure}

	The CMC was firstly identified as a coherent elongated cloud at a single distance by wide-field infrared extinction maps from
	2MASS \citep{2009ApJ...703...52L}.
	In the following observations toward the CMC, 
	pc-scale filaments are identified  in subregions \citep{2014A&A...567A..10L, 2019ApJ...877..114C, 2020A&A...642A..76Z}.
	In this work, as shown in Figures \ref{fig-lv} and \ref{fig-m0_thrcolor},
	a 72 pc long filament extends continuously in space and velocity and stretch across these subregions. 
	It has roughly the same length and width with "Nessie", the first identified large-scale massive filament, which
	has a length of about 80 pc and a width of 0.5 pc \citep{2010ApJ...719L.185J}. 
	Based on this physical size, the typical $M_{\rm line}$ of "Nessie"  estimated from HNC (1$-$0) fluxes is 
	$\sim$ 110 \msun\,pc$^{-1}$ \citep{2010ApJ...719L.185J}.  
	Nevertheless, \cite{2018A&A...616A..78M} estimated a mean $M_{\rm line}$ of $\sim$ 627 \msun\,pc$^{-1}$ with a physical 
	size of 67\,pc $\times$ 3\,pc inferred from the near infrared data of VISTA Variables in the 
	Via Lactea (VVV) survey and $Spitzer/GLIMPSE$ survey. 
	The "Nessie" Nebula contains a number of compact dense cores with a characteristic projected spacing of $\sim$ 4.5 pc.
	As shown in Figure \ref{fig-mline},  
	the $M_{\rm line}$ exceeds 100 \msun\,pc$^{-1}$ near the cluster LkH$\alpha$ 101, with its maximum reaches $\sim$ 330 \msun\,pc$^{-1}$ 
	($\sim$ 600 \msun\,pc$^{-1}$) at the projected radii of 0.6\,pc (2.4\,pc). 
	In contrast, the $M_{\rm line}$ is generally below $\sim$ 50 \msun\,pc$^{-1}$ ($\sim$ 100 \msun\,pc$^{-1}$) in other regions. 
	As the CO emission traces much diffuse ISM than the submillimeter dust continuum, 
	the $M_{\rm line}$ of the CMF may be underestimated by 2-3 times. 
	In contrast to "Nessie", the mass distribution of CMF is inhomogeneous. 
	In the traditional two-step model where thermally supercritical filaments form first and cores then form by 
	gravitational fragmentation \citep{1964ApJ...140.1056O,1992ApJ...388..392I,1997ApJ...480..681I}, 
	the high line mass region of the CMF maybe created by converging flows from the low mass regions.
	In a high shock velocity model \citep{2020arXiv201202205A}, the high line mass region can be created from the high density blobs in a short time, 
	so the parent molecular cloud of CMF may be  inhomogeneous.

	Because the mass distribution is roughly symmetrical within the $R_{\rm flat}$ 
	(see Figure \ref{fig-cum}), 
	we use 2\,$R_{\rm flat}$ (0.6 pc) as the width of the CMF to compare with the theoretical self-gravitating cylinder.
	The CMF has a higher mean values of the excitation temperature (10.4 $\pm$ 3.1 K)
	and the column density (3.64 $\pm$ 0.18 $\times$ 10$^{21}$ cm$^{-2}$) than the CMC.
	The mean $M_{\rm line}$ estimated  from the enclosed $M_{\rm line}$ profile (see Figure \ref{fig-cum}) is $\sim$ 26 \,\msun\,pc$^{-1}$,
	which is larger than the critical line mass 
	($M_{\rm line, crit} =$ 2\,$\upsilon^{2}/G \sim $ 464\,$(\Delta V)^{2}\approx 16.7 \,\mathrm{M}_{\odot} \mathrm{pc}^{-1}$)
	dominated by thermal support ($\Delta V \sim c_{s}\approx 0.19 \mathrm{~km} \mathrm{~s}^{-1}$) 
	and less than the critical line mass ($\sim$ 335.2\,\msun\,pc$^{-1}$)
	dominated by turbulent support ($\Delta V \sim \sigma_{tot} \approx$ 0.85 km\,s$^{-1}$) (see Figure \ref{fig-mline}).
	The detailed calculation method of the isothermal sound speed $c_{\mathrm{s}}$ and $\sigma_{tot}$
	was introduced in \cite{2017ApJ...838...49X,2019ApJ...880...88X}. 
	It indicates that the CMF is under the gravitationally unstable conditions and should fragment into dense cores,
	as we see in the \ei  images. In the process, turbulent pressure can support the filament against gravitational collapse.
	
\subsection{Kinematics of the CMF} \label{sec: kinematic of the filament}	
	
	 Systematic velocity gradients across the major filament have been observed in 
	 the pc-scale filamentary structures with low line mass ($\lesssim$ 100\,$\mathrm{M}_{\odot} \mathrm{pc}^{-1}$)
	 in the regions of Serpens South, 
	 Serpens Main, NGC\,1333 region  
	\citep{2018ApJ...853..169D}, the northwestern part of the L1495 \citep{2018PASJ...70...96A}, 
	 B211/B213 of the Taurus cloud \citep{2013A&A...550A..38P,2019A&A...623A..16S}, 
	and in the subregions of high line mass filaments ($\gtrsim$ 100\,$\mathrm{M}_{\odot} \mathrm{pc}^{-1}$) 
	 such as the CMC L1482 filament \citep{2021ApJ...908...86A} and the LBS 23 (HH24-HH26) region in Orion B \citep{2021ApJ...908...92H}.

	 The interpretation of the velocity gradient involves several different filament formation mechanisms, 
	 which are debated between internally-driven and externally-driven cloud and filament evolution.
	 The numerical simulations about externally-driven formation mechanisms include, e.g.  oblique MHD shock 
	 compression \citep{2015A&A...580A..49I,2018PASJ...70S..53I}, colliding flows \citep{2014ApJ...791..124G}, 
	 and direct compression of interstellar turbulence \citep{2014ApJ...785...69C}.
	The simulations from \cite{2014ApJ...791..124G} have demonstrated that filaments can be formed by colliding flows 
	via the gravitational collapse of a molecular cloud as a whole. 
	But such global contraction would not cause systematic velocity gradient across the major axis \citep{2020MNRAS.494.3675C}.
	\cite{2015A&A...580A..49I} and \cite{2018PASJ...70S..53I} have argued that multiple compressions associated with expanding bubbles 
	can create star-forming filamentary structures within sheet-like molecular cloud.
	Based on the MHD simulations, a similar model of anisotropic filament formation in shock-compressed layers has been proposed by 
	\cite{2014ApJ...785...69C}. According to these models, the velocity gradient is a projection effect of the accreting material 
	in a 2D flow within the dense layer created by colliding turbulent cells. 
	But such simulations reproduced filaments with lower masses/line-masses comparing with the CMF.
	In the simulation of \cite{2020arXiv201202205A}, high line-mass filaments of a few pc 
	can be created by gas flows driven by the high velocity shock wave ($v_{\mathrm{sh}} \simeq$ 10 $\mathrm{kms}^{-1}$) in a 
	timescale of the creation of dense compressed sheet-like region, or by the compressive flow component involved in the initial 
	turbulence at a slow shock velocity ($v_{\mathrm{sh}} \simeq$ 3 $\mathrm{kms}^{-1}$) case in a 
	relatively longer timescale ($\gtrsim$ 2\,Myr).
	However, the current simulations of converging flows induced by shock waves or turbulence at small scale do not support 
	the formation of such a large-scale filament as the CMF.

	Recently, \cite{2021ApJ...908...86A} analyze the  CMC L1482-S filament (NE1 in this work), which has a length of $\sim$ 0.9\,pc and 
	a mean $M_{\rm line}$ of $\sim$ 205\,\msun\,pc$^{-1}$ at the projected radii of 1\,pc  (measured from $Herschel$ data), using IRAM multi-tracer 
	($\mathrm{C}^{18} \mathrm{O}(1-0), \mathrm{N}_{2} \mathrm{H}^{+}(1-0), \mathrm{HCO}^{+} (1-0), \mathrm{HNC}(1-0)$). 
	The anti-symmetric of the radial velocity gradient of the inner portion 
	(1.43 km\,s$^{-1}$pc$^{-1}$ at  $r \lesssim$ 0.25 $\mathrm{pc}$, 
	0.17 km\,s$^{-1}$pc$^{-1}$ at 0.25 $\mathrm{pc}$  $\lesssim r \lesssim$ 0.4 $\mathrm{pc}$) 
	are interpreted by solid-body rotation,
	which is involved in the internally-driven filament formation mechanisms.
	Combing the "zig-zag" morphology and the flipped magnetic field,  
	\cite{2021ApJ...908...86A} suggested that rotating  "corkscrew" filaments would remove inner angular momentum via the magnetic 
	breaking, leading to collapse of the filament. 
	The magnetic field would transport this rotational energy from small scales near the  filament ridgeline to larger radii.
	\cite{2021ApJ...908...92H} observed a clear radial velocity gradient in the centern filament with a length of 0.16\,pc of
	 LBS 23 (HH24-HH26) region in Orion B though ALMA $\mathrm{N}_{2} \mathrm{D}^{+}$ observation.
	Based on the similarity between the angular momentum profile of filament and that of the dense cores or envelopes, 
	they suggest the angular momentum of the dense circumstellar environments may be linked to the rotation of small  filament. 
	The rotation may be acquired from ambient turbulence as they did not see any evidence of large-scale flows towards the 
	central  filament and radial velocity gradients in lower-density gas maps. 
	For the CMF, a systematic radial velocity gradient (0.82 km\,s$^{-1}$pc$^{-1}$) at larger radii ($\sim$ 1\,pc) can be seen in Figure \ref{fig-m1_L}.
	 Assuming that the rotating cylinder model  is suitable for the CMF and the velocity difference between the projected radii ($r$) and the ridgeline 
	 is the rotation velocity, 
	the angular frequency $(\omega)$ is $\sim 2.66 \times 10^{-14}\mathrm{~s}^{-1}$, which is  slightly lower than that 
	estimated from the inner portion ($r \lesssim$ 0.25 $\mathrm{pc}$) of 
	L1482-S filament \cite[$4.53 \times 10^{-14} \mathrm{~s}^{-1}, $][]{2021ApJ...908...86A}. 
	The two comparable $\omega$ indicate the radial gradients observed at low density and outer radii and the inner high density region may  
	correlate with each other.

	The interstellar environment where the CMF is located may offer us another possibility to explain
	the nature of the radial velocity gradient observed in the CMF.
	Through inspecting the spatial distribution of the H$\alpha$ and Planck 857 GHz emission in the Taurus-Perseus 
	region on scales up to $\ge$200 pc,  
	\cite{2019A&A...623A..16S} proposed an accretion model and suggested that the B211/B213 filament may have formed as a result of 
	compression of the Per OB2 association and the Local Bubble surrounding the Sun.
	The distances of the Per OB2 association, Taurus cloud, and the CMC are estimated to be 340 
	pc  \citep{1993BaltA...2..214C}, 140 pc \citep{1978ApJ...224..857E},
	and  470\,pc \citep{2019ApJ...879..125Z}, respectively.  
	As described in \cite{2008hsf1.book..308B}, 
	the most distant parts of the Taurus molecular cloud complex may be interacting with a supershell blown by the Per OB 2 association. 
	 In this picture, the Taurus cloud and the CMC may be on the front and back edge of the shell, respectively.
	The systematic velocity gradient across the CMF observed in this work is in a roughly opposite direction 
	 with respect to the B211/B213 filament \cite[see, e.g.][]{2019A&A...623A..16S}.
	This is consistent with the feature of the compression on the surrounding molecular clouds from the Per OB2 association.
	The wide-field atomic hydrogen and CO surveys towards the Per OB2 supershell and further detailed analysis about 
	the morphology, kinematics and timescale of the Taurus-Auriga, California, and Perseus molecular clouds are 
	presented in Chen et al. 2021. submited.
	They suggest that the large-scale converging supersonic flows  powered by the feedback from the Per OB2  association 
	may account for the CMF formation.

	  As shown in Figure \ref{fig-pv_acc},  along the major filament, the profile of $v_{\rm LSR}$ of \tht shows an oscillation pattern, 
	  which is suggested to be  related to core-forming flows 
	  \cite[see, e.g.][]{2011A&A...533A..34H, 2013A&A...554A..55H}.
	  All the \ei dense cores located within 0.6\,pc (5 beam size) along the ridgeline  are overlaid.   
	  Along the filament, we use a criterion similar to \citet{2018ApJ...853..169D} to estimated the velocity gradient,
	  that is, if the velocity difference is greater than $0.5 \mathrm{~km} \mathrm{~s}^{-1}$ (3 times velocity resolution) 
	  over a length of about 4 beam widths ($\sim$0.5 pc), the velocity gradient will be considered. 
	  Twenty-three segments are found to satisfy this criterion, their velocity gradients are estimated from the linear fitting (mpfit) of the 
	  m1 profile,  and their masses are estimated from \tht using the same method as in the calculation of the filament mass. 
	   Based on a simple cylindrical model from \cite{2013ApJ...766..115K}, the accretion rates could be estimated by
	  $\dot{M} = \frac{\bigtriangledown V_{\shortparallel }M}{tan(\alpha)}$, here the angle of the inclination to the plane 
	  of the sky  $\alpha$ is assumed to be 45\degr.  
	  The measured lengths,  velocity differences,  velocity gradients,  masses, and accretion rates of the 
	  23 segments are listed in Table \ref{tb-segment}. 
	  The mean length and mean velocity difference are $\sim$ 2.3 pc and $\sim$ 0.92 km\,s$^{-1}$, respectively.  
	  The accretion rates range from 9.1 to 144.1 M$_{\sun}$\,Myr$^{-1}$ with a mean value of 35.1 M$_{\sun}$\,Myr$^{-1}$.
	  
	  Around the LkH$\alpha$ 101 cluster ([19, 25] pc along the CMF), the mass accretion rates are estimated to be
	  at a magnitude of $\sim$ 101 M$_{\sun}$\,Myr$^{-1}$,
	  which is about five times larger than the mean value of other regions ($\sim$ 21 M$_{\sun}$\,Myr$^{-1}$). 
	  In fact, as mentioned above, the mass derived from \tht may be underestimated by 2-3 times, 
	  then the accretion rates here may also be underestimated.
	  The estimated accretion rate around the LkH$\alpha$ 101 cluster is comparable with the high mass star-forming regions,
	  e.g. Monoceros R2 \cite[the mean accretion rate  into the central hub $\sim$ 100\,M$_{\sun}$\,Myr$^{-1}$, ]
	  []{2019A&A...629A..81T}, 
	  Orion \cite[the accretion rate per molecular finger $\sim$ 55 M$_{\sun}$\,Myr$^{-1}$, ][]{2017A&A...602L...2H}. 
	 The mean value measured in the other regions is similar to that found in low mass star formation region,  e.g. 
	  Serpens South \cite[the accretion rate at a 0.33 pc long filament  $\sim$ 30 M$_{\sun}$\,Myr$^{-1}$; ][]{2013ApJ...766..115K}.
	 
          The positions of the \ei cores which hint density enhancements along the filament seem to be coupled to
          the oscillation patterns  of the profile of $v_{\rm LSR}$. 
          This indicates the oscillation patterns are likely formed via convergent flows caused by star formation. 
          The fragmentation characteristic length scale ($\sim$ 2.3 pc) is comparable with that found in "Nessie" ($\sim$ 4.5 pc).
          The theory of "sausage" instability \citep{2010ApJ...719L.185J} predicts that in an isothermal cylinders of finite radius $R_{\rm f}$ 
          surrounded by an external, uniform medium, the characteristic wavelength
	  $\lambda_{\rm{crit}}=22 H$ for $R_{\rm{f}} \gg H$,  while for $R_{\rm f} \ll H$,   $\lambda_{\rm{crit}}=11 H$,
	   where $H=c_{\rm{s}}\left(4 \pi G \rho_{\rm{c}}\right)^{-1 / 2}$ is the isothermal scale height.
	  The central mass density $\rho_{\rm{c}}$  along the axis of the cylinder is $\rho_{\rm{c}}=n_{\rm{c}} \mu m_{\rm{H}}$, 
	  where $n_{\mathrm{c}}=2 N_{\mathrm{c}} / \pi R_{\mathrm{flat}}$ \citep{2018MNRAS.478.2119L}. 
	  From the Plummer-like function fitting to the column density profile, 
	  $N_{\mathrm{c}} \sim$ 2.387 $\times$ 10$^{21}$ cm$^{-2}$, which results in  $\lambda_{\rm{crit}} \sim$ 1.6\,pc.  
	  In the case of turbulent pressure dominates over thermal pressure,  the effective scale height $H$ would become several times larger
	  ($\sim$ 0.35\,pc), then $\lambda_{\rm{crit}} \sim$ 3.9-7.8\,pc. 
	  The mean fragmentation wavelength ($\sim$ 2.3\,pc) along the CMF is comparable with the predictions of fragmentation 
	  of the self gravitating fluid cylinder.
	  %Using Faraday rotation measurements,  \cite{2018A&A...614A.100T} found that the magnetic fields ($\sim$ 
	  %several hundred $\mu \mathrm{G}$) around the CMF are pointing towards us at the south side and 
	  %pointing away from us at the north side.  They suggested that a helical magnetic field morphology is wrapping the filament.  
	  According to the fragmentation model of an isothermal self-gravitating gas 
	   cylinder including a helical magnetic field proposed by \citet{1993PASJ...45..551N} (see their equations 6 and 7), 
	  assuming the angle between magnetic field direction and main axis is 45$\degr$, 
	  the predicted strength of the undisturbed magnetic field along directions of azimuthal and the major axis 
	  are $\sim$ 10\,$\mu \mathrm{G}$ and $\sim$ 27\,$\mu \mathrm{G}$, respectively.  
	  %The magnetic fields  on scales of the whole filament may be supercritical. 
	  \cite{2016A&A...590A...2S} proposed a model of slingshot mechanism in which protostars are always undergoing 
	  transverse acceleration and will  "eject" from the dense filaments when they  becomes sufficiently massive.
	  They suggested  the observed undulations which indicates the transverse waves propagating through the filament 
	  are magnetically induced. As shown in  Figure \ref{fig-cores_2} and Figure \ref{fig-pv_acc}, 
	  the \ei cores  lie close to the ridge in position and velocity. 
	  %%The observed undulations in velocity indicate the gas are accelerating. 
	  According to the slingshot mechanism, the embedded protostar system may not be massive enough to decouple from the filament,  
	  thus the CMF may at an early stage characterized by supercritical magnetism and low star formation rates 
	  (compared to Orion region, see Section \ref{sec: star formation}) .

\subsection{Substructures} \label{sec: kinematic of fiber-like structures}

	Compared with the regions where the substructures are seen to overlap or roughly parallel to the major axis of the CMF (e.g. NE1, SW1), 
	the regions where the construction of the sub-structure exhibits the hub-like features seem to contains more YSOs or 
	more star formation activities (e.g. SE, NE2 and SW2).
	In NE1, 
	the velocity gradients of the substructures are consistent with the dynamics description 
	for the filament L1482 in \cite{2014A&A...567A..10L}, i.e., the filament might involve supersonic converging flows 
	to feed the stellar cluster LkH$\alpha$ 101.
	\cite{2021ApJ...908...86A} suggested the "zig-zag" morphology and velocity wiggles are consistent with a
	 helical velocity field in a filament with a corkscrew-like 3D morphology. 
	 The 3D structure in this segment is explained by rotational motion about the long axis.
	In SW1,  the wave-like velocity structures indicate accretion flows associated with cores formation.
	In the X-structure region, the distribution of the sub-structures is consistent with the dynamic analysis in \cite{2017ApJ...840..119I}. 
	As shown in the bottom panel of Figure \ref{fig-fiber-pv}, the spiny structures may be connected with the velocity gradients on beam size scale. 
	Though as described in \cite{2018MNRAS.479.1722C}, 
	there is very poor correspondence between the fibers in PPV space and the sub-filaments in PPP space, 
	both the two types of structures are considered to be driven from inhomogeneities in the turbulent accretion flow onto the main filament. 
	Similarly, we suggest that the substructures resolved in PP space and the spiny structure in PV plots may be connected with 
	the accretion flows along sub-structures onto the ridgeline of filament. 
		
	As shown in Figure \ref{fig-pv_ver}, there are more than one single velocity component across the CMF.
	In the panels 4, 5, and 6, there is a blueshifted velocity component at about -3 km\,s$^{-1}$ and a redshifted 
	velocity component at about 1 km\,s$^{-1}$.
	The two velocity components seem to separate on both sides of the CMF ($v_{\rm LSR} \sim$ -1 km\,s$^{-1}$), 
	which seems to be consistent with the segments shown in Figure \ref{fig-pvshow} (e.g. [35, 40] pc, [40, 50] pc along the CMF).  
	For the blueshifted velocity components, they all finally merge into the CMF. 
	For the redshifted velocity components, they are separated from or faintly surrounding the CMF (see the panels 4, 5, 6).
	The multiple velocity components look  like branch-like structures of the CMF.
	The multiple velocity components across the ridgeline of the filament may be connected with 
	several optional origins, such as rotation \citep{2019MNRAS.489.4771G}, cloud-envelope velocity difference, 
	infall, or cloud-cloud collision. 
		
\subsection{Early-Phase Star Formation Activities} \label{sec: star formation}
	 	
	 As described in \cite{2009ApJ...703...52L}, the star formation activity in the CMC is significantly less than the OMC. 
	 The physical origin of the different levels of star formation activity between the two clouds may be due to the 
	 different amount of high extinction (A$_{K} \textgreater$ 1 mag) material that the clouds contain. 
	 The CMC contains only $1 \,\%$ of its mass above A$_{K}$ = 1 mag, 
	 while the OMC contains roughly $10\,\%$ of its mass above the same extinction level \citep{2009ApJ...703...52L}.
	 As shown in Table \ref{tb-cloud}, 
	 the high density mass traced by C$^{18}$O accounts for about $3.5\,\%$ of the total mass traced by $^{12}$CO. 
	 In the MWISP survey toward the OMC, the mass calculated from the three isotopologue lines are 
	 about 4.9$\times$10$^{4}$\,\msun  \,($^{12}$CO), 
	 4.3$\times$10$^{4}$\,\msun ($^{13}$CO), 0.9$\times$10$^{4}$\,\msun  (C$^{18}$O), respectively (Ma et al. in prep). 
	 The high density (C$^{18}$O) mass accounts for about $18\,\%$ of the total mass ($^{12}$CO).
	 This ratio is significantly larger than that of the CMC. 
	 
\begin{figure}[htb]
		\centering
		\includegraphics[width=1.0\textwidth]{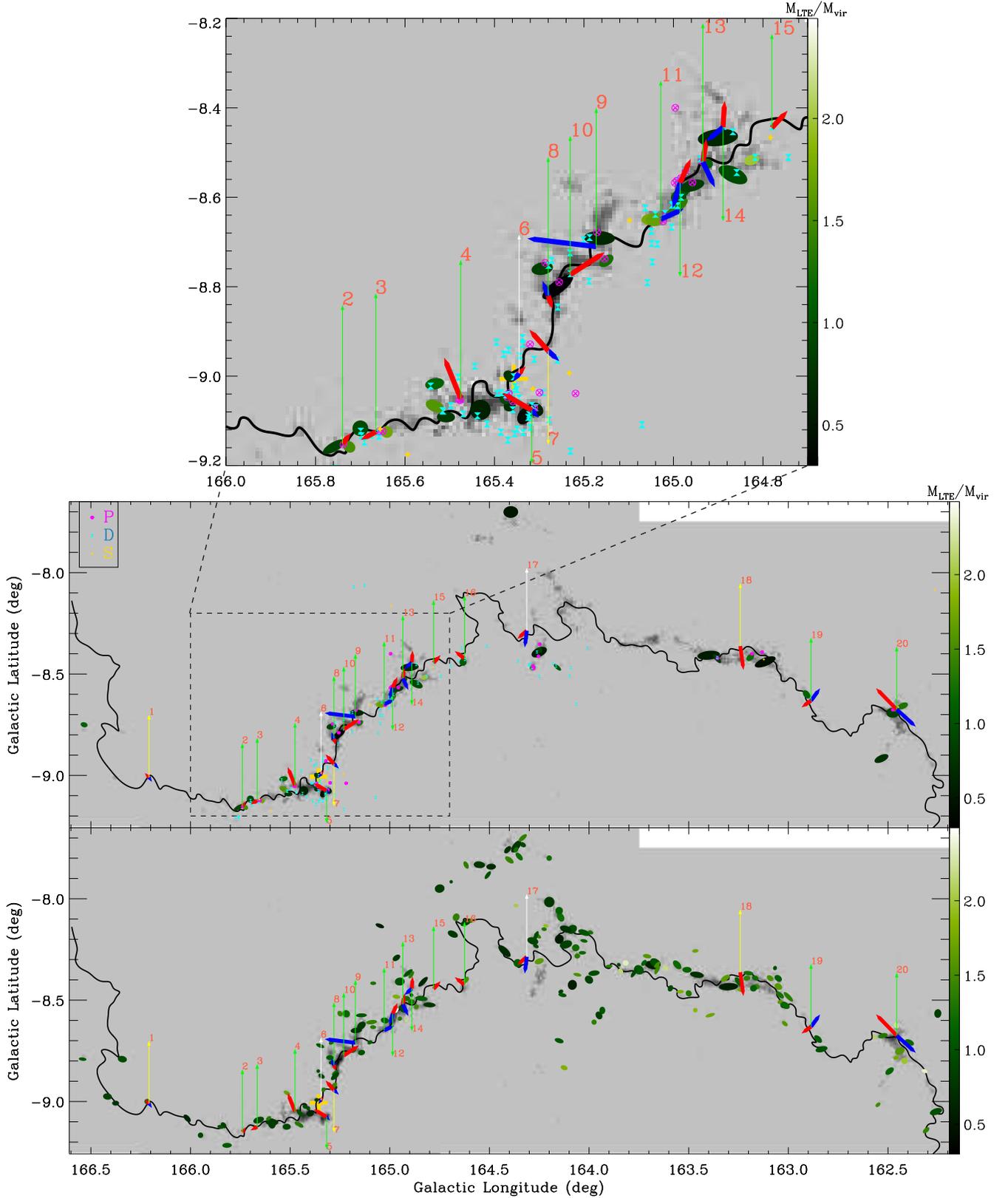}\\%\textit{(a)}
		\caption{Upper panel: The zoomed-in area of the dashed box in the middle panel.
		Middle panel: The distribution of the 83 dense cores (filled ellipses) that overlap with YSOs. 
		The dense cores are colorscaled  by $\alpha\rm_{vir}$.
		The blueshifted lobes and redshifted lobes of the  outflow candidates  are simply marked by the blue 
		and red arrows with corresponding lengths, respectively.
		The angle between the outflow candidate and the major axis of the CMF seems to have a random distribution.
		The green (yellow) arrows represent the 
		candidates may be associated  with the dense cores that overlap with YSOs
		(starless cores),  and white represents the opposite. 
		The YSOs are same as Figure\ref{fig-outflow}. 
		The grayscale background is the \ei m0 map.
		Bottom panel: Same as the middle panel but for the 181 starless cores with outflow candidates overlaid.
		 }
		\label{fig-cores_2}		
	\end{figure}

\begin{figure}[htb]
		\centering
		\includegraphics[width=1.0\textwidth]{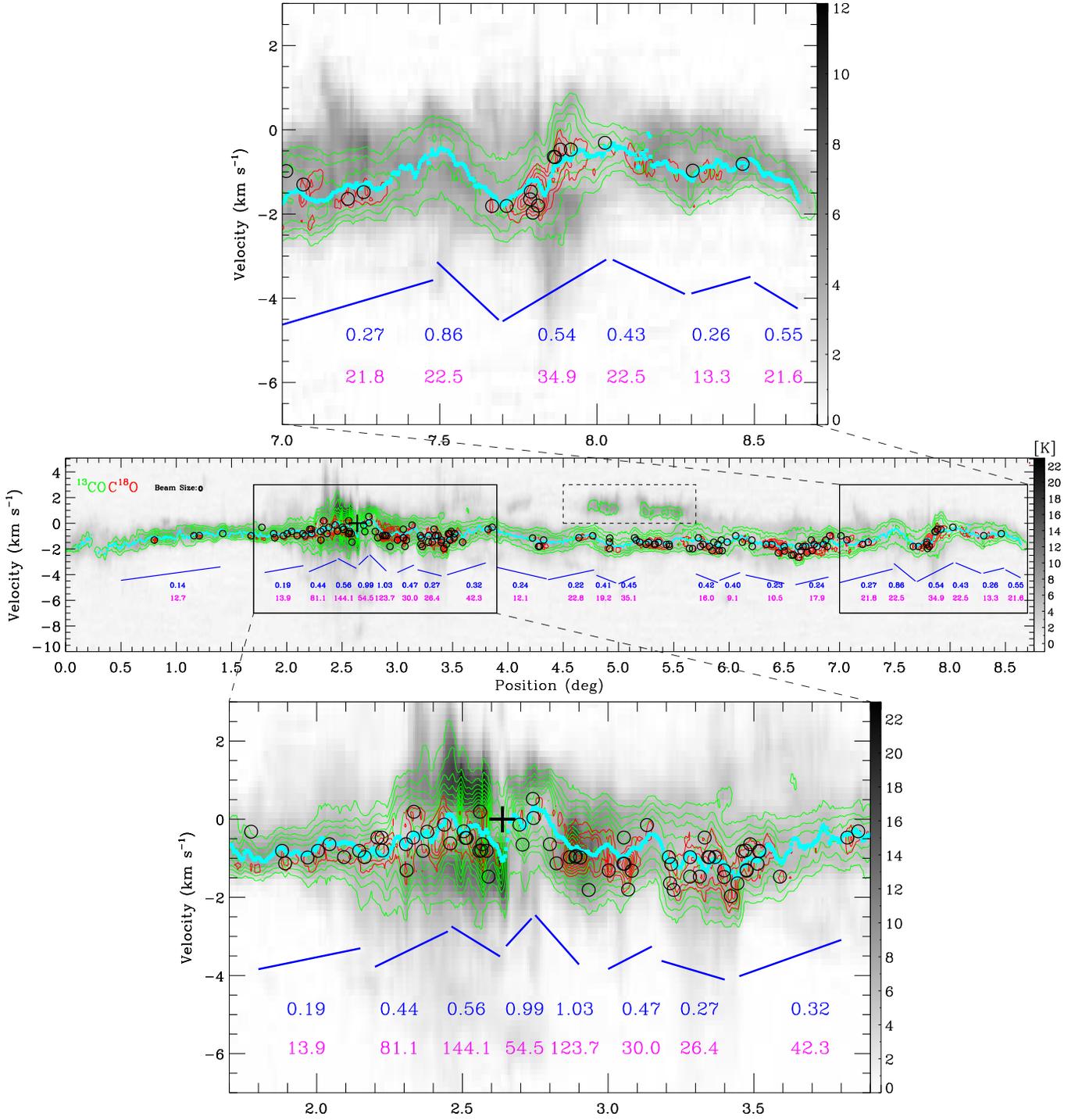}% ;angle=90
		\caption{Top panel: Same as the middle panel but for the zoomed-in area on the west end of the CMF. 
		Middle panel: The PV diagram along the ridgeline of the CMF
	     with a width of 8 pixels from east to west. The grayscale background,  green contours, and red contours represent 
	     the $^{12}$CO, $^{13}$CO, and C$^{18}$O emission, respectively.
	     The contours start from 3\,$\sigma$  to peak value in steps of 3\,$\sigma$ for \tht and 
	     from 2\,$\sigma$ to peak value in steps of 1\,$\sigma$ for \ei, respectively.   
	     The cyan curve indicates the m1 profile of $^{13}$CO.
	     In order to ensure the monotonicity of velocity component, the \tht emission in the black dashed box is neglected 
	     in the m1 estimation.
	     The blue lines indicate the linear fitted results with corresponding velocity gradients (blue)
	      in the unit of km\,s$^{-1}$pc$^{-1}$ and accretion rates (magenta)  in the unit of M$_{\sun}$\,Myr$^{-1}$ 
	      marked below.
	     The black circles represent the extracted \ei dense cores. %(see, Section \ref{sec: Prestellar cores}). 
	     The solid line box marks the zoomed-in area. 
	     The colorbar is in the unit of K. 
	     Bottom panel: Same as the middle panel but for the zoomed-in area around the LkH$\alpha$ 101 cluster. 
	      }	   
	\label{fig-pv_acc}
	\end{figure}

	 From the PV diagram along the CMF, we have identified 20 outflow candidates with 31 individual lobes. 
	 Among them, the potential driving positions of 13 candidates are overlapped with the YSOs.
	 As shown in Figure \ref{fig-outflow}, the lobes overlap with each other.
	 The ambiguity might be due to the interactions between the outflows, or the outflows are intrinsically weak. 
	 The investigation of outflow-filament correlation \cite[e.g. position angle between outflows and filaments; see, e.g.,][]{2019ApJ...874..104K}
	 is helpful to search for evidence of the accretion-filament correlation. 	 
	 For the CMF, nevertheless, see, e.g., Figure \ref{fig-cores_2}, the angle between the outflow candidate and the major axis 
	 of the CMF seems to have a random distribution.
	 For a comparison, through the high spatial resolution ($\sim$ 8\,$\arcsec$) observations of CO (1-0) toward 
	 the Orion A GMC \citep{2020ApJ...896...11F}, 45 protostellar outflows with 67 individual lobes were identified 
	 from an area of approximately 2 deg$^{2}$. 
	 The detected number per square area in the CMC  ($\sim$ 2 deg$^{-2}$)
	 is significantly less than that in the OMC ($\sim$ 20 deg$^{-2}$). 
	 On the one hand, it may be due to our insufficient resolution that cannot fully distinguish the outflow.
	 On the other hand, it implies that stellar activity in the CMC is far less active than OMC. 
	 In addition, the line detection rate along the CMF is $\sim$ 3.5 deg$^{-1}$ in region E 
	 and 0.9 deg$^{-1}$ in region W, respectively.
	 The western filament around L1478 contains less star formation activity and maybe in a relatively earlier 
	 stage than the eastern one around L1482.

        To check the star formation activity, we present the spatial distributions of the \ei cores  
	 that overlap with YSOs and outflow candidates in Figure \ref{fig-cores_2}. 
	 Of these identified dense cores, $53\,\%\,(120/225)$ are distributed along the ridgeline of the CMF within a distance of 
	 0.3 pc  and  $74\,\%\,(167/225)$ locate within a distance of 0.6 pc. 
	 This result is consistent with that of $Herschel$ PACS observations \citep{2010A&A...518L..95H}.
	 Of the 181 starless dense cores, 94\,\% (171/181) have $\alpha\rm_{vir}$ less than 2 and 37\,\% (67/181) 
	 have $\alpha\rm_{vir}$ less than 1.
	 The 67 starless dense cores with $\alpha\rm_{vir}$ less than 1 may have potential to further collapse to form protostars, 
	 which could be regarded as prestellar cores.  
	 The free collapse timescale $t_{\mathrm{ff}}=\left(\frac{3 \pi}{32 G\rho}\right)^{1/2}$ ($\rho$ is the 
	 median value of the number density of the gravitationally bound starless cores and is $\sim6.3\times 10^{3} \rm cm^{-3}$) 
	 is estimated to be about 0.4 Myr. 
	 The total mass of the bound starless cores ($M_{\rm starless, bound}$) is $\sim$ 328\,\msun. 
	 Assuming the star formation efficiency per free-fall time ($\frac{\epsilon}{f}$) is $\sim$ 1.8\,\% \citep{2010ApJ...724..687L},  
	 the star formation rate (SFR $\sim \frac{\epsilon}{f} \times M_{\rm starless, bound}/t_{\mathrm{ff}}$) is $\sim$ 
	 152 $\mathrm{M}_\odot \mathrm{Myr}^{-1}$. 
	 The present SFR estimated from YSO population by 
	 $\mathrm{SFR}=0.25 N(\mathrm{YSOs}) \times 10^{-6} \rm M_{\odot} \mathrm{yr}^{-1}$ \citep{2010ApJ...724..687L}
	 is $\sim$ 44 $\mathrm{M}_\odot \mathrm{Myr}^{-1}$.  The sample indicates the SFR of California may increase in the future.

	  As shown in Figure \ref{fig-cores_2}, the average distance to the CMF ridgeline is $\sim$ 0.57 pc for 
	  the group of dense cores that overlap with YSOs (upper panel) 
	  and $\sim$ 0.88 pc for the group of starless cores (bottom panel), respectively. 
	  The positional correlation between cores and the ridgeline is weaker for starless cores than for the dense cores that overlap with YSOs.
	  This may imply that star formation occurs primarily along the major filament \citep{2013ASPC..476...95A} and then begins outside. 
	  Among the 20 outflow candidates, 15 candidates may be 
	  associated with the dense cores with YSOs (as marked by green arrows) and 3 may be associated with starless cores (as marked by yellow arrows).  
	  Thus, most (18/20) outflow candidates we identified in this work are associated with dense cores or YSOs and 
	  can be considered relatively reliable.

\section{Conclusions} \label{sec: conclusions}		 
	
	Based on the wide-field CO (J=1-0) isotopologue lines observations  with the  PMO-13.7 m millimeter telescope, 
	we investigate the physical properties of the molecular gas in the CMC region.
	The main results are summarized as follows: 
	
	1, According to the space and velocity coherence, we find a large-scale filamentary molecular cloud (named as CMF in this work)
	with a length of about 72 pc from the \tht emission. 
	The mean line mass within the width of 0.6\,pc of the CMF is estimated to be 26\,\msun\,pc$^{-1}$.
	Comparing with the critical line masses,  the CMF is under the gravitationally unstable conditions and fragments into dense cores. 
	
	2, The CMF is found to show systematic velocity gradients perpendicular to the major axis. 
	By checking the PV diagrams across the CMF, the typical systematic velocity gradient is estimated to be $\sim$ 0.82 km\,s$^{-1}$pc$^{-1}$.
	The potential origin including filament rotation and compression by large-scale converging flows driven by stellar feedback from the Per OB2 association are discussed.

	3, Along the CMF, the large-scale velocity profile shows an oscillation pattern with 
	a mean wavelength of $\sim$ 2.3 pc, which appears to follow the predictions of the "sausage" instability theory.
	The oscillatory pattern is consistent with the kinematic model of core-forming motions, and  
	the accretion rates are estimated to be about $\sim$ 101 M$_{\sun}$\,Myr$^{-1}$ around the LkH$\alpha$ 101 cluster, 
	and $\sim$ 21 M$_{\sun}$\,Myr$^{-1}$ in other regions. 
	The magnitudes of the two rates are in agreement with the estimated rates of high and low star formation in other studies, respectively.
	
	4, Combing the \ei channel maps and the DisPerSE algorithm, substructures with slight velocity difference 
	along the CMF are resolved. There are multiple velocity components displaying branch-like structures 
	in some segments of the CMF. The dynamic of the substructures and the multiple velocity components along the CMF may be 
	connected with the formation scenario of converging flows for filaments.
	
	5, We identify a total of 225 \ei dense cores with a mean size of 0.12 pc and a mean mass of 3.3\,\msun\,in the CMC region.
	Of the 225 dense cores, 181 are found to have no association with YSOs. 
	Of the 181 starless dense cores, $94\,\% \,(171/181)$ have $\alpha_{\text{vir}}$ less than 2 and $37\,\% \,(67/181)$ 
	have $\alpha_{\text{vir}}$ less than 1.
	We find 20 outflow candidates with 31 individual lobes  along the CMF. 
	The detection rate of the outflow candidates in eastern CMF  is higher than that in western CMF.
	Combining the distributions of \ei dense cores, YSOs and outflow candidates, we find there are 18 candidates
	may be related to \ei dense cores or YSOs, which can be considered relatively reliable.
	Based on these results,  intense early phase star formation activity is revealed in the CMC region.

\begin{acknowledgments}	
	We thank the anonymous referee for the helpful suggestions to improve our work.
	This work was supported by the National Key R\&D Program of China, grant NO. 2017YFA0402702, 
	and the National Natural Science Foundation of China, grants NO. 12041305, 11629302 and 11933011.
	This research made use of the data from the Milky Way Imaging Scroll Painting (MWISP) project, which is a multi-line survey in ${ }^{12} \mathrm{CO} /{ }^{13} \mathrm{CO} / \mathrm{C}^{18} \mathrm{O}$ along the northern galactic plane with PMO-13.7m telescope. We are grateful to all the members of the MWISP working group, particularly the staff members at PMO-13.7m telescope, for their long-term support. MWISP was sponsored by National Key R\&D Program of China with grant 2017YFA0402701 and  CAS Key Research Program of Frontier Sciences with grant QYZDJ-SSW-SLH047.	
\end{acknowledgments}	
		
%\appendix

%\section{Appendix information}
	
\clearpage	
	
%\begin{longrotatetable}
\begin{deluxetable*}{ccccccccccccc}
		%\tabletypesize{\scriptsize} 
		%\tablecolumns{14} 
		\setlength{\tabcolsep}{0.030in}
		\tablecaption{Properties of the identified dense cores \label{tb-clumps}}
		\tablewidth{0pt}
		\tablehead{ \colhead{Cores}   & \colhead{$ \Theta_\mathrm{maj} $} & \colhead{$ \Theta_\mathrm{min} $} & \colhead{$PA$} 
		& \colhead{$v_{\rm LSR}$} & \colhead{$\Delta V$} & \colhead{$R\rm_{eff}$\tablenotemark{a}} 
		& \colhead{$T\rm_{ex}$} & \colhead{$ N \rm_{H_{2}} $} 
		& \colhead{$ n\rm_{H_{2}} $}  & \colhead{$ M_\mathrm{LTE} $}  & \colhead{$ M_\mathrm{vir}$}  & \colhead{$ \alpha_\mathrm{vir}$}\\
		& \colhead{($ \arcmin $)} & \colhead{($ \arcmin $)} & \colhead{($\degr$)} 
		& \colhead{(km\,s$^{-1}$)}  &\colhead{(km\,s$^{-1}$)} & \colhead{(pc)} 
		& \colhead{(K)}   & \colhead{($ 10^{21}$cm$^{-2} $)} 
		& \colhead{($ 10^{3}$cm$^{-3}$)}  & \colhead{(M$_{\sun} $)}  & \colhead{(M$_{\sun}$)}  & \\
		\colhead{$(1)$} & \colhead{$(2)$} & \colhead{$(3)$} & \colhead{$(4)$} & \colhead{$(5)$} & \colhead{$(6)$} 
		& \colhead{$(7)$}  & \colhead{$(8)$} & \colhead{$(9)$} & \colhead{$(10)$} 
		& \colhead{$(11)$} & \colhead{$(12)$} & \colhead{$(13)$} 
		}
                \startdata
MWISP G165.259-8.801    &     1.73     &     4.98     &     53.20     &        -0.94     &      0.63      &  0.19        &        20.72    &        13.11    &      26.98           &      53.1      &     15.70    &    0.30   \\
MWISP G165.333-9.089    &     1.75     &     2.48     &     33.20     &        -0.80     &      0.67      &  0.13        &        30.54    &        14.03    &      49.32           &      28.6      &     11.81    &    0.41   \\
MWISP G165.432-9.075    &     2.73     &     2.72     &     0.00      &       -0.62      &     0.65       &  0.17        &       16.88     &       7.64      &      17.38           &      26.8      &     15.40    &    0.57   \\
MWISP G163.682-8.327    &     4.59     &     2.35     &     12.40     &        -2.13     &      0.44      &  0.21        &        7.59     &       2.96      &      5.24            &      15.1      &     8.69     &    0.58   \\
MWISP G165.165-8.691    &     1.86     &     4.12     &     91.80     &        -1.14     &      0.57      &  0.18        &        11.56    &        5.24     &      11.68           &      18.9      &     12.04    &    0.64   \\
MWISP G162.475-8.691    &     3.96     &     2.93     &     -1.60     &        -1.47     &      0.69      &  0.22        &        8.14     &       4.88      &      8.25            &      26.6      &     22.21    &    0.83   \\
MWISP G164.984-8.591    &     1.46     &     2.90     &     55.60     &        -1.47     &      0.57      &  0.12        &        11.34    &        4.23     &      15.12           &      8.5       &     8.45     &    1.00   \\
MWISP G164.925-8.524    &     1.34     &     1.63     &     107.10    &         -1.97    &       0.62     &  0.08        &         12.47   &         5.18    &      40.88           &      5.3       &     6.16     &    1.15   \\
MWISP G164.149-8.200    &     2.61     &     2.76     &     31.40     &        -1.49     &      0.39      &  0.17        &        11.04    &        3.09     &      7.20            &      10.5      &     5.44     &    0.52   \\
MWISP G163.400-8.407    &     6.80     &     2.60     &     -5.90     &        -1.65     &      0.55      &  0.28        &        9.08     &       3.94      &      5.16            &      32.9      &     17.71    &    0.54   \\
		\enddata
		\tablenotetext{a}{Deconvolution Radii: $ R_{eff} = \frac{1}{2}\times d \times \frac{\pi}{2} \sqrt{\Theta_\mathrm{maj}\Theta_\mathrm{min}-\theta_\mathrm{beam}^2} $.}
		%\tablenotetext{b}{Equivalent radius: $r_\mathrm{eq} = d\sqrt{\Theta_\mathrm{maj}\Theta_\mathrm{min}-\theta_\mathrm{beam}^2}/\sqrt{2\ln2}$, where $ \theta_\mathrm{beam} = 25\arcsec $.}
		%	\tablenotetext{a}{Intensity weighted velocity with respect to the systemic velocity.}
		%	\tablenotetext{b}{Derived from the distance of peak of each lobe to the MIR point source.}
		\tablecomments{Table \ref{tb-clumps} is published in its entirety in the machine-readable format. A portion is shown here for guidance regarding its form and content.}
\end{deluxetable*}
%\end{longrotatetable}    

		\begin{deluxetable}{cccccccc}
		\tabletypesize{\small} 
%		\tablecolumns{5} 
		\setlength{\tabcolsep}{0.15in}
		\tablecaption{Median values of the properties of the dense cores \label{tb-comparecores}}
		\tablewidth{0pt}
		\tablehead{ \colhead{dense cores} & \colhead{$R_{\rm eff}$} & \colhead{$T\rm_{ex}$} & \colhead{$N\rm_{H_{2}}$} & 
			\colhead{$n\rm_{H_{2}}$} & \colhead{$M\rm_{LTE}$}& \colhead{$M\rm_{vir}$}   & \colhead{$\alpha\rm_{vir}$} \\
			& \colhead{(pc)} & \colhead{(K)} & \colhead{($ 10^{21} $ cm$^{-2}$)} & \colhead{($ 10^{3} $ cm$^{-3}$)}  
			& \colhead{(\msun)}  & \colhead{(\msun)}  & } 
		\colnumbers
		\startdata
		all the dense cores     &  $0.11^{+0.08}_{-0.06}$     &  $9.8 ^{+9.7}_{-2.3}$        &  $1.69^{+3.4}_{-0.8}$          &   $7.9^{+36.7}_{-4.8}$       & $3.0^{+10.6}_{-2.0}$       & $3.4^{+5.8}_{-2.3}$      &  $1.11^{+0.85}_{-0.53}$ \\
		cores with YSOs          & $ 0.12 ^{+0.13}_{-0.04}$    &  $10.6 ^{+19.8}_{-2.6}$      &  $2.41^{+6.3}_{-1.2}$          &   $8.8^{+35.0}_{-5.5}$        & $5.6^{+22.8}_{-4.3}$      & $5.8^{+9.9}_{-4.3}$      &  $0.93^{+0.79}_{-0.48}$ \\
		starless  cores            &  $ 0.10 ^{+0.07}_{-0.05}$    &  $ 9.4 ^{+5.4}_{-1.9}$        &  $1.60^{+1.9}_{-0.7}$           &   $7.7^{+37.0}_{-4.6}$        & $2.8^{+6.3}_{-1.8}$          & $3.2^{+4.6}_{-2.1}$       &  $1.13^{+0.90}_{-0.46}$ \\
		\enddata 
		%\tablenotetext{b}{FWHM.}
	\end{deluxetable}

%	\startlongtable
        \begin{longrotatetable}
	\begin{deluxetable*}{lllccccccccclllll}
		%\tabletypesize{\scriptsize} 
		%\tablecolumns{14} 
		\setlength{\tabcolsep}{0.06in}
		\tablecaption{Identification of the outflow candidates \label{tb-outflow}}
		\tablewidth{0pt}
		\tablehead{ \colhead{NO.} & \colhead{$l_{\rm zoom}$} & \colhead{$b_{\rm zoom}$} &  \colhead{$z_{\rm zoom}$} & \colhead{$v\rm_{center}$} & \colhead{$v\rm_{range}$(blue)} & \colhead{$v\rm_{range}$(red)} 
		& \colhead{ $n \times \sigma\rm_{blue}$}  & \colhead{$n \times \sigma\rm_{red}$}  & \colhead{ID}  & \colhead{$l$} & \colhead{$b$} & \colhead{$l_{\rm red}$} & \colhead{$b_{\rm red}$} & \colhead{$l_{\rm blue}$} & \colhead{$b_{\rm blue}$}  & YSOs\\
		& \colhead{($\degr$)} & \colhead{($\degr$)} & \colhead{($\degr$)} & \colhead{(km\,s$^{-1}$)} & \colhead{(km\,s$^{-1}$)} & \colhead{(km\,s$^{-1}$)} & \colhead{} 
		& \colhead{} & & \colhead{($\degr$)} & \colhead{($\degr$)} & \colhead{($\degr$)} & \colhead{($\degr$)} & \colhead{($\degr$)} & \colhead{($\degr$)} & \colhead{(P, D, S)} \\
		\colhead{$(1)$} & \colhead{$(2)$} & \colhead{$(3)$} & \colhead{$(4)$} & \colhead{$(5)$} & \colhead{$(6)$} & \colhead{$(7)$}  & \colhead{$(8)$}& \colhead{$(9)$}
		& \colhead{$(10)$} & \colhead{$(11)$}  & \colhead{$(12)$}& \colhead{$(13)$} & \colhead{$(10)$} & \colhead{$(11)$}  & \colhead{$(12)$}& \colhead{$(13)$}}
                \startdata
(1) & 166.202 & -9.000 & 0.08 & 0 & [ -2 , 0] & [ 1 , 2] & 33 & 10 & 1 & 166.208 & -9.010 & 166.229 & -8.988 & 166.193 & -9.032 & \\ 
(2) & 165.710 & -9.146 & 0.15 & 0 & [ -3 , -2] & [ 1 , 3] & 14 & 14 & 2 & 165.740 & -9.151 & 165.721 & -9.127 & nan & nan & P \\ 
 & & & & & & & & & 3 & 165.665. & -9.125 & 165.699 & -9.142 & nan & nan & P or D \\ 
(3) & 165.493 & -9.030 & 0.30 & -1 & [ -6 , -3] & [ 3 , 6] & 25 & 10 & 4 & 165.475. & -9.050 & 165.512 & -8.959 & nan & nan & P \\ 
(4) & 165.320 & -9.080 & 0.11 & -1 & [ -7 , -2] & [ 2.2, 5] & 40 & 9 & 5 & 165.317 & -9.076 & 165.382 & -9.038 & 165.300 & -9.093 & P or D \\ 
(5) & 165.350 & -8.990 & 0.06 & -1 & [ -7 , -2] & [ 1 , 3] & 35 & 47 & 6 & 165.345 & -8.993 & 165.330 & -8.978 & 165.359 & -9.013 & \\ 
(6) & 165.285 & -8.930 & 0.11 & -1 & [ -7 , -2] & [ 1 , 2] & 35 & 45 & 7 & 165.280 & -8.944 & 165.322 & -8.896 & 165.255 & -8.966 & \\ 
(7) & 165.280 & -8.820 & 0.08 & -1 & [ -4 , -2] & [ 1.5, 3] & 16 & 9 & 8 & 165.280 & -8.820 & 165.270 & -8.850 & 165.293 & -8.787 & \\ 
(8) & 165.210 & -8.700 & 0.30 & -1.5 & [ -4 , -3] & [ 1 , 3] & 8 & 13 & 9 & 165.173 & -8.710 & nan & nan & 165.326 & -8.692 & D \\ 
 & & & & & & & & & 10 & 165.231 & -8.773 & 165.154 & -8.723 & nan & nan & D \\ 
(9) & 164.960 & -8.570 & 0.22 & -2 & [ -7 , -5] & [ 2 , 3] & 5 & 8 & 11 & 165.028 & -8.649 & nan & nan & 164.977 & -8.623 & P or D \\ 
 & & & & & & & & & 12 & 164.985 & -8.568 & 164.962 & -8.514 & 165.002 & -8.620 & P or D \\ 
 & & & & & & & & & 13 & 164.934 & -8.521 & 164.927 & -8.470 & 164.907 & -8.578 & P or D \\ 
(10) & 164.887 & -8.430 & 0.15 & -1.2 & [ -6 , -4] & [ 2 , 3] & 4 & 13 & 14 & 164.889 & -8.444 & 164.886 & -8.384 & 164.927 & -8.475 & D \\ 
(11) & 164.700 & -8.428 & 0.30 & -1 & [ -4 , -2] & [ 0 , 1] & 24 & 30 & 15 & 164.780 & -8.445 & 164.745 & -8.406 & nan & nan & D \\ 
 & & & & & & & & & 16 & 164.625 & -8.423 & 164.674 & -8.387 & nan & nan & D \\ 
(12) & 164.325 & -8.312 & 0.16 & -2 & [ -4 , -3] & [ 2 , 3] & 13 & 15 & 17 & 164.313 & -8.286 & 164.360 & -8.327 & 164.323 & -8.369 & \\ 
(13) & 164.243 & -8.400 & 0.20 & -1.5 & [ -4 , -3] & [ 0 , 2] & 25 & 15 & 18 & 163.241 & -8.362 & 163.225 & -8.480 & nan & nan & \\ 
(14) & 162.900 & -8.635 & 0.20 & 0 & [ -3 , -2] & [ 1 , 2] & 15 & 5 & 19 & 162.886 & -8.631 & 162.934 & -8.665 & 162.841 & -8.571 & \\ 
(15) & 162.448 & -8.692 & 0.30 & -1 & [ -5 , -3] & [ 0 , 2] & 15 & 5 & 20 & 162.456 & -8.675 & 162.560 & -8.570 & 162.360 & -8.760 & P or D \\ 
		\enddata
		%\tablenotetext{a}{.}
		%\tablenotetext{b}{Equivalent radius: $r_\mathrm{eq} = d\sqrt{\Theta_\mathrm{maj}\Theta_\mathrm{min}-\theta_\mathrm{beam}^2}/\sqrt{2\ln2}$, where $ \theta_\mathrm{beam} = 25\arcsec $.}
		%	\tablenotetext{a}{Intensity weighted velocity with respect to the systemic velocity.}
		%	\tablenotetext{b}{Derived from the distance of peak of each lobe to the MIR point source.}
	\end{deluxetable*}
	\end{longrotatetable}  

	\begin{deluxetable}{lccccc}
		\tabletypesize{\small}
		\setlength{\tabcolsep}{0.15in}
		\tablecaption{Properties of the segments along the major filament \label{tb-segment}}
		\tablewidth{0pt} 
		%\tablenum{5}
		\tablehead{ \colhead{Segment} & \colhead{Length} & \colhead{Velocity difference} & \colhead{Velocity gradient} & \colhead{Mass} 
		& \colhead{Accretion rate}\\
		\colhead{(pc, pc)}&\colhead{(pc)} &\colhead{(km\,s$^{-1}$)} & \colhead{(km\,s$^{-1}$\,pc$^{-1}$)} & \colhead{(\msun)} & \colhead{(M$_{\sun}$\,Myr$^{-1}$)} 
		} 
		\colnumbers
		\startdata                                                              
(4.1  ,     11.5)    &    7.4     &     1.45     &       0.14    &    87.4      &    12.7      \\  
(14.8 ,     17.6)    &    2.9     &     0.61     &       0.19    &    73.2      &    13.9      \\  
(18.0 ,     20.1)    &    2.1     &     1.08     &       0.44    &    179.6     &    81.1      \\  
(20.2 ,     21.6)    &    1.4     &     1.03     &       0.56    &    251.5     &    144.1     \\  
(21.7 ,     22.5)    &    0.7     &     0.96     &       0.99    &    54.1      &    54.5      \\  
(22.6 ,     23.8)    &    1.2     &     1.07     &       1.03    &    118.0     &    123.7     \\  
(24.6 ,     25.8)    &    1.2     &     0.62     &       0.47    &    62.6      &    30.0      \\  
(26.1 ,     27.9)    &    1.8     &     0.69     &       0.27    &    95.8      &    26.4      \\  
(28.3 ,     31.2)    &    2.9     &     1.21     &       0.32    &    128.0     &    42.3      \\  
(32.0 ,     35.7)    &    3.7     &     1.09     &       0.24    &    48.6      &    12.1      \\  
(35.8 ,     39.2)    &    3.4     &     0.69     &       0.22    &    99.6      &    22.8      \\  
(39.4 ,     40.9)    &    1.5     &     0.64     &       0.41    &    45.4      &    19.2      \\  
(41.0 ,     42.2)    &    1.2     &     0.61     &       0.45    &    76.4      &    35.1      \\  
(46.8 ,     48.4)    &    1.6     &     0.6      &       0.42    &    37.8      &    16.0      \\  
(48.5 ,     50.0)    &    1.5     &     0.78     &       0.40    &    22.4      &    9.1       \\  
(50.4 ,     53.3)    &    2.9     &     0.95     &       0.23    &    45.1      &    10.5      \\  
(54.1 ,     56.6)    &    2.5     &     0.73     &       0.24    &    72.4      &    17.9      \\  
(57.4 ,     61.4)    &    3.9     &     1.27     &       0.27    &    79.5      &    21.8      \\  
(61.4 ,     63.1)    &    1.6     &     1.3      &       0.86    &    25.7      &    22.5      \\  
(63.2 ,     65.9)    &    2.7     &     1.42     &       0.54    &    63.0      &    34.9      \\  
(66.0 ,     67.9)    &    1.9     &     0.99     &       0.43    &    50.7      &    22.5      \\  
(68.1 ,     69.6)    &    1.5     &     0.51     &       0.26    &    50.5      &    13.3      \\  
(69.7 ,     70.9)    &    1.1     &     0.95     &       0.55    &    38.4      &    21.6      \\      
		\enddata
	\end{deluxetable}

\end{document}